\documentclass[11pt]{article}

\usepackage[margin=.75in]{geometry}
\usepackage{datetime}
\usepackage{amsmath, amsthm, amssymb,fancyhdr,mathrsfs,  enumerate}
\usepackage{latexsym}
\usepackage[round, authoryear, comma, sort&compress]{natbib}
\usepackage{hyperref}
\usepackage[utf8]{inputenc}
\usepackage{graphicx,float}
\usepackage[export]{adjustbox}

\newtheorem{lemma}{Lemma}

\renewcommand{\P}{\mathbb{P}}

\newcommand{\E}{\mathbb{E}}
\newcommand{\I}{\mathbb{I}}
\newcommand{\R}{\mathbb{R}}

\begin{document}

\begin{center}
      \Large{\bf {Multiple Observers Ranked Set Samples for Shrinkage Estimators}}
\end{center}
\begin{center}
 \noindent{{\sc Andrew David Pearce} and  
 {\sc Armin Hatefi} {\footnote{Corresponding author:\\ Email: ahatefi@mun.ca and Tel: +1 (709) 864-8416} } }

\vspace{0.5cm}
\noindent{\footnotesize{\it Department of Mathematics and Statistics, Memorial University of Newfoundland, St. John's, NL, Canada. \\
}}
\end{center}

\begin{center} {\small \bf Abstract}:
\end{center}
{\small
Ranked set sampling (RSS) is used as a powerful data collection technique for situations where 
measuring the study variable requires a costly and/or tedious process while the sampling units can
 be ranked easily (e.g., osteoporosis research). In this paper, we develop ridge and Liu-type 
 shrinkage estimators under RSS data from multiple observers to handle the collinearity problem 
 in estimating coefficients of linear regression, stochastic restricted regression and logistic 
 regression. Through extensive numerical studies, we show that shrinkage methods with the multi-observer 
 RSS result in more efficient coefficient estimates. The developed methods are finally applied to bone mineral data for 
analysis of bone disorder status of women aged 50 and older.
}


\section{ Introduction } \label{sec:intro}
In many medical surveys (e.g., osteoporosis research), measuring the response variable (e.g., 
diagnosis osteoporosis status) requires a costly and tedious process. Despite this challenge, 
practitioners typically have access to many easily attainable concomitants (e.g., demographic 
and laboratory characteristics). In these situations, the research question is how to incorporate
 these easy-to-measure characteristics into data collection and obtain more representative samples from the population.  

As a bone metabolic disorder, osteoporosis happens when the density of bone structure of the body reduces significantly.
 This significant reduction leads to various major health problems, including susceptibility to skeletal fragility 
 and osteoporotic fractures in the hip and femoral area \citep{black1992axial,cummings1995risk}. About 33\% of women 
 and 20\% of men experience an osteoporotic fracture in the age group of 50 and older \citep{melton1998bone}. More 
 than 50\% of individuals who experienced an osteoporotic hip fracture will no longer be able to live independently and 
about 28\% of them will die within one year  from  complications of the fractured bone \citep{bliuc2009mortality,neuburger2015impact}. 
As the aged population increases in many countries, one should anticipate a rise in the prevalence of osteoporosis-related health problems. 

The expert panel of WHO introduces the Bone Mineral Density (BMD) as the most reliable factor for bone disorder
 analysis and diagnosis. Despite this reliability, measuring BMD is costly and tedious. BMD measurements are 
 taken from dual X-ray absorptiometry (DXA) imaging. Once images are taken, medical experts are needed for manual
  segmentation of images and extracting necessary measurements.  While measuring BMD is difficult, clinicians 
  typically have access to easily attainable characteristics about patients such as weight, BMI, age and BMD scores
   from earlier surveys. We believe ranked set sampling, as a cost-effective technique, can be used to 
   incorporate these characteristics into data collection and results in more efficient estimates for the 
   parameters of the osteoporosis population.  

To construct an RSS (i.e., ranked set sampling/sample) from the osteoporosis population,  first $H^2$ patients are 
identified and allocated to $H$ sets of size $H$. Patients are then ranked (without measuring 
their BMD) using an inexpensive characteristic (e.g., age). We select the individual with rank $r$ from 
the $r$-th set ($r=1,\ldots,H$) and measure her BMD score. This results in an RSS of size $H$. The cycle is then repeated 
$n$ times to find a balanced RSS sample of size $N=nH$. RSS has found applications in a wide variety of fields such as breast cancer \citep{hatefi2017improved,zamanzade2018proportion}, 
osteoporosis \citep{nahhas2002ranked},  education \citep{wang2016using}, environments 
\citep{amiri2014resampling,ozturk2014estimation,frey2012nonparametric}, fishery \citep{hatefi2015mixture},
 trauma \citep{helu2011nonparametric} to name a few. 
  
Collinearity is one of the most common issues in linear regression and logistic regression. In the presence of 
collinearity, the least squares estimates become unreliable and lead us to misleading results. 
Despite a rich literature of RSS data for linear regression and logistics regression\citep{lynne1977ranked,muttlak1995parameters,zamanzade2018proportion,chen2005ranked,hatefi2020efficient,alvandi2021estimation}, 
there are very few works investigated the RSS for shrinkage estimates to cope with the collinearity issue. 
Although \cite{ebegil2021some} recently proposed ridge and Liu-type estimators with median RSS for the 
collinearity, the proposed shrinkage estimators are only developed for the linear regression model (i.e.,
 continuous response). In addition, the developed RSS estimators can only use a single characteristic for 
 ranking involved in RSS. Plus, the methods are not capable of using the ties information in RSS data.  

Despite the importance of logistic regression in medical studies, to our best knowledge, there is no 
research in literature investigated the use of RSS data for shrinkage methods to deal with the collinearity 
issue in logistic regression and stochastic restricted regression models.
While \cite{zamanzade2018proportion,chen2005ranked,hatefi2020efficient,alvandi2021estimation} used RSS data for
 analysis of logistic regression and binary data, the research question here differs from them. In this 
 manuscript, we shall employ RSS data to develop shrinkage estimators for the coefficients of the logistic regression. Unlike 
\cite{chen2005ranked}, the estimation methods here do not require training data.  
 While \cite{zamanzade2018proportion,chen2005ranked,alvandi2021estimation} focus on logistic regression where 
 the population proportion 
 is fixed, here we treat the response prevalence $\pi({\bf x})$ in a general form (i.e., as the 
 generalized linear 
 function of predictors) so that it changes from one individual to another and is of course subject to collinearity issue.  
 In this manuscript, we use multi-observer RSS of \cite{alvandi2021estimation} that inherits the tie flexibility of
 \cite{frey2012nonparametric} and multiple observers of \citep{ozturk2013combining} in data collection. 
 We then develop ridge and Liu type estimators under multi-observer RSS to overcome the collinearity issue in linear regression, stochastic restricted regression and logistic regression. 
 The developed methods are finally applied to bone mineral data for analysis of bone disorder status of women aged 50 and older.
  
 This manuscript is organized as follows. Section \ref{sec:rss} describes the construction of RSS data from multiple observers. 
 The development of shrinkage methods with SRS and RSS data are discussed in Sections \ref{sec:srs} and \ref{sec:est_rss}. 
 The estimation methods are evaluated through various simulation studies and real data examples in Sections \ref{sec:num}
 and  \ref{sec:real}. Summary and concluding remarks are finally presented in Section \ref{sec:sum}.   
 
\section{Multi-observer RSS}\label{sec:rss}
\cite{frey2012nonparametric} introduced an RSS method that is able to incorporate as many tied 
units as desired into data collection. Despite this flexibility, this RSS scheme
can only use a single observer for ranking in RSS. 
 \cite{alvandi2021estimation} proposed an RSS scheme that simultaneously enjoys the tie information of 
\citep{frey2012nonparametric} and multiple observers of 
\citep{ozturk2013combining}. In this manuscript, we investigate the properties of multi-observer RSS 
data of \citep{alvandi2021estimation} to deal with the collinearity issue.

Let $({\bf X},{\bf y})$ denote a multivariate random variable where ${\bf y}$ and ${\bf X}=({\bf x}_1,\ldots,{\bf x}_p) $ 
denote the response vector and the design matrix with $p$ predictors, respectively.  
Suppose ${\bf R}=(R_1,\ldots,R_K)$ is the vector of $K$ easy-to-measure external concomitant 
variables which are used for ranking purposes in RSS. 

In this manuscript, we focus on balanced multi-observer RSS (MRS) data. 
 Let $H$ and $n$ denote set and cycle sizes of the scheme, respectively.
 To collect MRS data of size $N=nH$, similar to the standard RSS (described in Section \ref{sec:intro}), 
 we plan to measure the $r$-th judgemental order statistic $Y_{[r]j}$ 
  from the $r$-th set in the $j$-th cycle.
  Let ${\bf U}^{[r]}_{j}=(u^{[r]}_{1,j},\ldots,u^{[r]}_{H,j})$ represent the $H$ units in the $r$-th set. 
  We rank the units with respect to their corresponding concomitant variables 
  ${\bf R}^{[r]}_{k,j}=({R}^{[r]}_{1,k,j},\ldots,{R}^{[r]}_{H,k,j})$ for $k=1,\ldots,K$.
  Suppose ${\bf W}^{[r]}_{k,j}$ denotes the weight matrix (of size $H \times H$) that records the ranks and 
  ties information assigned by observer ${\bf R}^{[r]}_{k,j}$.
    \cite{frey2012nonparametric} introduced the Discrete Perceived size   (DPS) model to implement the tie in RSS. 
    The DPS model gives a tied rank to units $u^{[r]}_{l_1,j}$ and $u^{[r]}_{l_2,j}$ if $[u^{[r]}_{l_1,j}/c]=[u^{[r]}_{l_2,j}/c]$,
     where $[\cdot]$ indicates the floor function and $c$ is a user-chosen parameter. 
  If unit $u^{[r]}_{l_1,j}$ receives rank 
  $l_2$ by observer ${\bf R}^{[r]}_{k,j}$, the $(l_1,l_2)$ entry of ${\bf W}^{[r]}_{k,j}$ 
  becomes one; otherwise the entry will be zero.
   Once we recorded weight matrices ${\bf W}^{[r]}_{k,j}$ for $k=1,\ldots,K$, the ranking information is combined by 
   \[
   {\bar{\bf W}}^{[r]}_{j} = \sum_{k=1}^{K} \eta_k {\bf W}^{[r]}_{k,j}, ~~~~~~~ \eta_k=\frac{|\rho_k|}{\sum_{k'=1}^{K} |\rho_{k'}|}
   \]
   where $\rho_k$  denotes the correlation between $R_k$ and ${\bf y}$. The unit with the  
   highest weight in the $r$-th column of ${\bar{\bf W}}^{[r]}_{j}$ is then selected and measured  with response 
   $Y_{[r]j}$  and $p$ predictors ${\bf X}_{[r]j}=(x_{1[r]j},\ldots,x_{p[r]j})$. In a similar fashion, we measure 
   other MRS statistics. Eventually, the MRS data (of size $N=nH$) is given by
$\{ (Y_{[r]j},{\bf X}_{[r]j},{\bar w}^{[r]}_{j}); r=1,\ldots,H;j=1,\ldots,n\}$.

In the multi-observer RSS scheme of \citep{alvandi2021estimation}, one can measure different  order 
statistics from different sets. 
While this proposal gives us information from all aspects of the population,   measuring only medians from all 
sets may be more beneficial when the population is symmetric. To this end, here we generalize the multi-observer 
RSS (MRS) of  \citep{alvandi2021estimation} to median multi-observer RSS (MMR) scheme where we  always measure
 medians from ${\bf U}^{[r]}_{j}$ sets for $r=1,\ldots,H$. Accordingly, the MMR data of size $nH$ is given by
 $\{ (Y_{[l]j},{\bf X}_{[l]j},{\bar w}^{[l]}_{j}); j=1,\ldots,nH\}$ where $l=(H+1)/2$ when $H$ is odd and 
 when $H$ is even, we use  $H/2$ for the first $N/2$ observations and $H/2 + 1$ for the other half.
For more details about the milt-observer RSS and median RSS scheme, see 
\citep{alvandi2021estimation,samawi2001estimation,ozturk2013combining} and references therein.

\section{Estimation Methods with SRS} \label{sec:srs}

Multiple linear regression model has found applications in many fields. Suppose the model is given by
\begin{align}\label{reg}
{\bf y} = {\bf X} {\bf \beta} + {\bf \epsilon},
\end{align}
where $\bf \beta$ represents the unknown coefficients of the model, ${\bf y}$ is $(N\times 1)$ vector of 
responses and  ${\bf X}$ represents non-random  $(N\times p)$ design matrix of $p$  explanatory variables 
$({\bf x}_1,\ldots,{\bf x}_p)$ with $\text{rank}({\bf X})=p < N$. Assume 
${\bf \epsilon} \sim N({\bf 0},\sigma^2 \I_N)$, with
 $\E({\bf \epsilon}) = {\bf 0}$ and $\text{Var}({\bf \epsilon})=\sigma^2 \I_N$ where $\sigma^2$ is the common 
 variance of error and $\I_N$ is $N$ dimensional identity matrix. Let ${\bf S} = {\bf X}^\top {\bf X}$ henceforth.
  The regression model \eqref{reg} can also be viewed in a canonical form. Using an
 orthogonal matrix $U$, one can diagonalize ${\bf S}$ by $ U^\top {\bf S} U = \Lambda$ where
 $\Lambda=\text{diag}(\lambda_1,\ldots,\lambda_p)$ with $\lambda_1 > \ldots > \lambda_p$ are the eigenvalues of ${\bf S}$.
In the canonical form, the regression model \eqref{reg} becomes 
$
{\bf y} = {\bf Z} {\bf \gamma} + {\bf \epsilon},
$
where ${\bf Z}={\bf X}^\top U$ and ${\bf \gamma}= U^\top {\bf \beta}$. 

 Least square method is the most common method to estimate the coefficients of the regression model \eqref{reg}.
The Least Squares (LS) estimator of ${\bf\beta}$ is then given by 
 \begin{align}\label{ls}
{\widehat{\bf \beta}}_{LS} = {\bf S}^{-1} {\bf X}^\top {\bf y}.
\end{align}
One big challenge of ${\widehat{\bf \beta}}_{LS}$ occurs in the presence of
collinearity where the explanatory variables are linearly dependent.  The Ridge method is 
considered as a solution to the problem. 
The ridge estimator ${\widehat{\bf \beta}}_{R}$ is given by 
 \begin{align}\label{ridge}
{\widehat{\bf \beta}}_{R} = ({\bf S} + k \I)^{-1} {\bf X}^\top {\bf y},
\end{align}
where $k$ represents the ridge parameter.
When the collinearity arises, 
the matrix ${\bf S}$ is then considered ill-conditioned; hence 
${\bf S}^{-1}$ becomes practically singular. The ridge method 
suggests to add $k$ to the diagonal elements of the ill-conditioned matrix ${\bf S}$ 
at the price of introducing a bias in the estimation procedure. 
When the collinearity is more sever, the ridge parameter 
alone may not be enough to overcome the ill-conditioned matrix. 
\cite{liu2003using} proposes the Liu-type  method as follows
 \begin{align}\label{liu}
{\widehat{\bf \beta}}_{LT} = ({\bf S} + k \I)^{-1} ({\bf X}^\top {\bf y}+ d {\widehat{\bf \beta}}_{R}),
\end{align}
where $k>0$, $d \in \R$ and ${\widehat{\bf \beta}}_{R}$ is given by \eqref{ridge}.


\subsection{Stochastic Restricted Estimators}\label{sub:sto}
In many situations, in addition to the regression model \eqref{reg}, one has prior information about the 
coefficients of the model in a form of a set of $j$ independent stochastic restrictions as 
\begin{align}\label{reg_sto}
{\bf r} = {\bf R} {\bf \beta} + {\bf e},
\end{align}
where ${\bf r}$ is a $(j \times 1)$  vector known responses, ${\bf R}$ is a $(j \times p)$ 
known matrix and ${\bf e}$ is a random vector with $\E({\bf e})={\bf 0}$, $\text{Var}({\bf e})=\sigma^2 {\bf\Omega}$ where ${\bf\Omega}$ 
is assumed to be a known semi-positive matrix. Assume that  ${\bf e}$ and ${\bf \epsilon}$ are stochastically independent. 
For more information about the regression model with stochastic restrictions, see \citep{yang2009improvement,li2010new}.
Using the stochastic restrictions \eqref{reg_sto}, \cite{theil1961pure} introduced a weighted mixed estimator ${\widehat{\bf \beta}}_{\text{ME}}$ 
 as
 \begin{align}\label{mixed}
{\widehat{\bf \beta}}_{\text{ME}} = ({\bf S} + v  {\bf R}^\top {\bf \Omega}^{-1} {\bf R})^{-1} 
({\bf X}^\top {\bf y}+ v {\bf R}^\top {\bf \Omega}^{-1} {\bf r}),
\end{align}
where $v$ is a non-stochastic scalar weight $0 < v \le 1$.
In the presence of collinearity,   \cite{hubert2006improvement} employed one parameter Liu estimate of \citep{kejian1993new} 
in the context of stochastic restricted regression. Hence, they proposed Mixed Liu estimator of ${\bf \beta}$ as follows
 \begin{align}\label{mixed_liu}
{\widehat{\bf \beta}}_{\text{MXL}} = ({\bf S} + \I)^{-1} ({\bf S} + d \I) {\widehat{\bf \beta}}_{\text{ME}},
\end{align}
where ${\widehat{\bf \beta}}_{\text{ME}}$ is given by \eqref{mixed} and $d \in \R$.
 \cite{yang2009alternative} replaced ${\widehat{\bf \beta}}_{LS}$ with ${\widehat{\bf \beta}}_{LT}$ 
in mixed estimation method. Accordingly, they proposed an alternative stochastic restricted Liu estimator as follows
 \begin{align}\label{srl}
{\widehat{\bf \beta}}_{\text{SRL}} = {\widehat{\bf \beta}}_{LT1} 
+ v {\bf S}^{-1} {\bf R}^\top ({\bf \Omega} + v {\bf R} {\bf S}^{-1} {\bf R}^\top)^{-1}
 ({\bf r}- {\bf R} {\widehat{\bf \beta}}_{LT1}). 
\end{align}
where ${\widehat{\bf \beta}}_{\text{LT1}}$ of \citep{kejian1993new} is given by
\begin{align}\label{liu_one}
{\widehat{\bf \beta}}_{\text{LT1}} = ({\bf S} + \I)^{-1} ({\bf S}+ d \I) {\widehat{\bf \beta}}_{LS}. 
\end{align}
As another method to handle the  collinearity, \cite{li2010new} used the properties of the ridge 
estimator in the context of stochastic restricted regression and defined mixed ridge estimator as follows
 \begin{align}\label{mixed_ridge}
{\widehat{\bf \beta}}_{\text{MXL}} = (\I + k {\bf S}^{-1})^{-1} {\widehat{\bf \beta}}_{\text{ME}},
\end{align}
where ${\widehat{\bf \beta}}_{\text{ME}}$ is given by \eqref{mixed}.  \cite{arumairajan2014improvement} employed
 ${\widehat{\bf \beta}}_{\text{R}}$ of \eqref{ridge} in mixed estimation method \eqref{mixed}  to deal with collinearity. Hence, the stochastic 
 restricted ridge estimator ${\widehat{\bf \beta}}_{\text{SRR}}$ is given by
 \begin{align}\label{srr}
{\widehat{\bf \beta}}_{\text{SRR}} = {\widehat{\bf \beta}}_{R} 
+ v {\bf S}^{-1} {\bf R}^\top ({\bf \Omega} + v {\bf R} {\bf S}^{-1} {\bf R}^\top)^{-1}
 ({\bf r}- {\bf R} {\widehat{\bf \beta}}_{R}). 
\end{align}
\subsection{Logistic Regression Estimators with SRS}\label{sub:log}
Logistic regression model plays a key role in analysis of binary responses. 
Let logistic model follow
 \begin{align}\label{logistic}
 \P(y_i=1 | {\bf X}) = g({\bf x}_i; {\bf \beta}) = 
 1/(1+\exp( -{\bf x}_i^\top {\bf \beta})),
\end{align}
where $\bf \beta$ represents the coefficients of the model, ${\bf y}$ is vector of 
binary responses, $g$ as the link function and ${\bf X}$ is non-random  $(n\times p)$ design 
matrix of $p$  explanatory variables 
$({\bf x}_1,\ldots,{\bf x}_p)$ with $\text{rank}({\bf X})=p < N$. 

The Maximum likelihood (ML) method is the common method to estimate the coefficients of logistic model \eqref{logistic}. 
 The log-likelihood function of ${\bf \beta}$ is given by 
 \begin{align}\label{ll-logistic}
 l({\beta}) = \sum_{i=1}^{n} - \log \left( 1+\exp( - {\bf x}_i^\top {\bf \beta}) \right)
 + y_i ( {\bf x}_i^\top {\bf \beta}).
 \end{align}
 In the absence of a closed-form maximum of \eqref{ll-logistic}, 
 one can find the ML estimate ${\widehat{\bf\beta}_{\text{ML}}}$ using Newton Raphson (NR) method. To do so,
  given ${\bf \beta}^{(l)}$ the estimate from $l$-th iteration, we iteratively update ${\bf \beta}^{(l+1)}$ as follows
   \begin{align}\label{nr}
 {\bf \beta}^{(l+1)} = {\bf \beta}^{(l)} - {\bf H}^{-1}\left(l({\bf \beta}^{(l)})\right)  {\nabla}_{\beta} l({\bf \beta}^{(l)}),
 \end{align}
 where ${\bf H}^{-1}\left(l({\bf \beta}^{(l)})\right)$ and ${\nabla}_{\beta} l({\bf \beta}^{(l)})$ are the Hessian matrix 
 and  the gradient 
  evaluated at ${\bf \beta}^{(l)}$, respectively. 
 
 The ${\widehat{\bf\beta}_{\text{ML}}}$  
 is also severely affected  when there exists a collinearity problem in the logistic regression. 
 To deal with the collinearity, \cite{schaefer1984ridge} introduced a ridge estimator as follows
   \begin{align}\label{ridge_log} 
{\widehat{\bf \beta}}_{\text{R,log}} 
= ( {\bf X}^\top {\bf V} {\bf X} + k \I)^{-1} {\bf X}^\top {\bf V} {\bf X} {\widehat{\bf\beta}_{\text{ML}}}.
\end{align}
When collinearity is severe, ${\widehat{\bf \beta}}_{\text{R,log}}$ may not be able to 
cope with the ill-conditioned matrix. \cite{inan2013liu} 
proposed the Liu-type logistic estimator with $k >0$, $d \in \R$ as follows 
  \begin{align}\label{liu_log} 
{\widehat{\bf \beta}}_{\text{LT,log}}  =
 ( {\bf X}^\top {\bf V} {\bf X} + k \I)^{-1} ({\bf X}^\top {\bf V} {\bf X} - d \I) {\widehat{\bf\beta}_{ML}}.
\end{align}

To obtain the ridge parameter $k$, there are three common choices for $k$ 
 \citep{schaefer1984ridge,inan2013liu}. They include 
$1/{\widehat{\bf\beta}}_{ML}^\top {\widehat{\bf\beta}_{ML}}, p/{\widehat{\bf\beta}}_{ML}^\top 
{\widehat{\bf\beta}_{ML}},(p+1)/{\widehat{\bf\beta}}_{ML}^\top {\widehat{\bf\beta}_{ML}}$ where $p$ 
denotes the number of predictors in the model. In this manuscript, 
we select $k$ using ${\widehat k}=(p+1)/{\widehat{\bf\beta}}_{ML}^\top {\widehat{\bf\beta}_{ML}}$. 
To determine $d$, \cite{inan2013liu} proposed a numerical approach to find 
the optimal $d$ value which minimizes  the mean square errors (MSE) of ${\widehat{\bf \beta}}_{\text{LT,log}}$.

\section{Estimation Methods with RSS}\label{sec:est_rss}
In this section, we shall use multi-observer RSS (MRS) to develop shrinkage estimators 
 for the parameters of  regression model \eqref{reg}, stochastic restricted model 
\eqref{reg_sto} and logistic regression model \eqref{logistic}. Although \cite{ebegil2021some} 
recently develop ridge and Liu-type estimators based on median RSS data for regression model \eqref{reg}, their method is 
not capable of combining ranking information from multiple sources in data collection. 
In addition, to our best knowledge, there is no research work  investigating 
the  RSS for shrinkage  estimation in stochastic restricted regression model  
and logistic regression in the presence of colinearilty.

We first investigate the estimation of regression model \eqref{reg}  based on MRS data. 
As described in Section \ref{sec:rss}, with set size $H$ and cycle size $n$,   
let  ${\bf X}_{\text{MRS}}$ be $(n H\times p)$ design matrix and  ${\bf y}_{\text{MRS}}$ be $(n H \times 1)$ response vector 
based on MRS data  when we ignore the weights. Let ${\bf S}_{\text{MRS}} = {\bf X}_{\text{MRS}}^\top {\bf X}_{\text{MRS}}$ henceforth.
 Based on  MRS data, the least square estimator of ${\bf \beta}$ is given by
 \begin{align}\label{ls_rss}
{\widehat{\bf \beta}}_{\text{LS,MRS}} = {\bf S}_{\text{MRS}}^{-1} {\bf X}_{\text{MRS}}^\top {\bf y}_{\text{MRS}}.
\end{align}
\begin{lemma}\label{moments_ls_rss}
$
\E({\widehat{\bf \beta}}_{\text{LS,MRS}}) =  {\bf \beta}
$ and
$
\text{Var} ({\widehat{\bf \beta}}_{\text{LS,MRS}}) = \sigma^2 {\bf S}_{\text{MRS}}^{-1}.
$
\end{lemma}
While ${\widehat{\bf \beta}}_{\text{LS,MRS}}$ is an unbiased estimator for ${\bf \beta}$, 
it is severely 
influenced by collinearity issue. Following 
\citep{schaefer1984ridge,ebegil2021some},  
one possible solution is to develop the ridge estimator of ${\bf \beta}$. Hence, the ridge estimator based on 
 MRS data is given by
 \begin{align}\label{ridge_rss}
{\widehat{\bf \beta}}_{\text{R,MRS}} = ({\bf S}_{\text{MRS}} + k \I)^{-1} {\bf X}_{\text{MRS}}^\top {\bf y}_{\text{MRS}}.
\end{align}
\begin{lemma}\label{moments_ridge_rss}
The expected value and covariance matrix of ${\widehat{\bf \beta}}_{\text{R,MRS}}$ are given by
$
\E({\widehat{\bf \beta}}_{\text{R,MRS}}) =  {\bf W}_{\text{MRS}} {\bf \beta}$ and
$ \text{Var} ({\widehat{\bf \beta}}_{\text{R,MRS}}) = \sigma^2 {\bf W}_{\text{MRS}} {\bf S}_{\text{MRS}}^{-1}  {\bf W}_{\text{MRS}}^\top
$
where ${\bf W}_{\text{MRS}} = (\I +k {\bf S}_{\text{MRS}}^{-1})^{-1}$.
\end{lemma}
Note that ${\widehat{\bf \beta}}_{\text{R,MRS}}$ is obtained by least square solution to 
the model ${\bf y}_{\text{MRS}} = {\bf X}_{\text{MRS}} {\bf \beta} + {\bf \epsilon}$ subject to 
${\bf 0}= k^{\frac{1}{2}} {\bf \beta} + \epsilon^\top$. As the ridge parameter increases,
 Lemma \ref{moments_ridge_rss} shows that the bias of ${\widehat{\bf \beta}}_{\text{R,MMR}}$ increases. 
 For this reason, it is advantageous to choose small $k$ in the ridge estimate. 
When $k$ is small, $({\bf S}_{\text{MMR}} + k \I)$ may still be ill-conditioned. 
Hence, the Liu-type estimator of ${\bf \beta}$ 
based on MRS data is given by:
  \begin{align}\label{liu_rss}
{\widehat{\bf \beta}}_{\text{LT,MRS}} = ({\bf S}_{\text{MRS}} + k \I)^{-1} ({\bf X}_{\text{MRS}}^\top 
{\bf y}_{\text{MRS}}+ d {\widehat{\bf \beta}}_{\text{R,MRS}}),
\end{align}
where $k>0$, $d \in \R$ and ${\widehat{\bf \beta}}_{\text{R,MRS}}$ is given by \eqref{ridge_rss}.
\begin{lemma}\label{moments_liu_rss}
The expected value and covariance of ${\widehat{\bf \beta}}_{\text{LT,MRS}}$ are given by
$
\E({\widehat{\bf \beta}}_{\text{LT,MRS}}) =  {\bf A}_{\text{LT,MRS}} {\bf \beta}
$ and
$
\text{Var} ({\widehat{\bf \beta}}_{\text{LT,MRS}}) = \sigma^2 {\bf A}_{\text{LT,MRS}} 
{\bf S}_{\text{MRS}}^{-1}  {\bf A}_{\text{LT,MRS}}^\top
$
where ${\bf A}_{\text{LT,MRS}}=({\bf S}_{\text{MRS}} + k \I)^{-1} \left( \I + d 
({\bf S}_{\text{MRS}}+k \I \right)^{-1}) {\bf S}_{\text{MRS}}$.
\end{lemma}
\subsection{Stochastic Restricted Regression with RSS}\label{sub:sto_rss}
Here, we employ the MRS data to estimate ${\bf \beta}$ 
in stochastic restricted regression model \eqref{reg_sto}. We first require the following Lemma (whose proof can be found in  \cite{rao1995linear}). 
\begin{lemma} \label{lemma_rao}
Let $A$ and $C$ be nonsingular matrices and $B$ and $D$ be matrices of proper orders. Then
$
(A + B C D)^{-1} = A^{-1} - A^{-1} B (C^{-1} + C A^{-1} B) D A^{-1}.
$  
\end{lemma}
\begin{lemma} \label{mixed_ls_rss}
Under the assumption of regression model \eqref{reg_sto} and Lemma \ref{lemma_rao},
\begin{itemize}
\item[i)] the weighted mixed estimator from MRS data is given by
 \begin{align*}\label{mixed_rss}
{\widehat{\bf \beta}}_{\text{ME,MRS}} = ({\bf S}_{\text{MRS}} + v  {\bf R}^\top {\bf \Omega}^{-1} {\bf R})^{-1} 
({\bf X}_{\text{MRS}}^\top {\bf y}_{\text{MRS}}+ v {\bf R}^\top {\bf \Omega}^{-1} {\bf r}),
\end{align*}
where $v$ is a non-stochastic scalar weight $0 < v \le 1$.
\item[ii)] ${\widehat{\bf\beta}}_{\text{ME,MRS}}$ can be written as a function of ${\widehat{\bf \beta}}_{\text{LS,MRS}}$ as
\[
{\widehat{\bf \beta}}_{\text{ME,MRS}} = {\widehat{\bf \beta}}_{\text{LS,MRS}} 
+ v {\bf S}_{\text{MRS}} ^{-1} {\bf R}^\top ({\bf \Omega} + v {\bf R} {\bf S}_{\text{MRS}}^{-1} {\bf R}^\top)^{-1}
 ({\bf r}- {\bf R} {\widehat{\bf \beta}}_{\text{LS,MRS}} ). 
\] 
\end{itemize}
\end{lemma}
The performance of ${\widehat{\bf \beta}}_{\text{ME,MMR}}$, as a function of  ${\widehat{\bf \beta}}_{\text{LS,MMR}}$, 
is severely affected with collinearity.
To overcome the problem, the first solution can be to incorporate the method of \cite{hubert2006improvement} 
into  ${\widehat{\bf \beta}}_{\text{ME,MMR}}$. Using Lemma \ref{mixed_ls_rss} and 
 \eqref{liu_one}, we propose the mixed Liu with MRS data as follows
 \begin{align}\label{mixed_liu_rss}
{\widehat{\bf \beta}}_{\text{MXL,MRS}} = ({\bf S}_{\text{MRS}} + \I)^{-1} ({\bf S}_{\text{MRS}} + d \I) {\widehat{\bf \beta}}_{\text{ME,MRS}},
\end{align}
where  $d \in \R$. Another method to attack the collinearity problem in 
${\widehat{\bf \beta}}_{\text{ME,MMR}}$ is to use Lemma \ref{mixed_ls_rss} and directly incorporate 
${\widehat{\bf \beta}}_{\text{LT1,MRS}}$  into all aspects of the estimator. 
Then, we can propose an alternative stochastic restricted Liu estimator based on MRS data as follows
 \begin{align}\label{srl_rss}
{\widehat{\bf \beta}}_{\text{SRL,MRS}} = {\widehat{\bf \beta}}_{\text{LT1,MRS}} 
+ v {\bf S}_{\text{MRS}}^{-1} {\bf R}^\top ({\bf \Omega} + v {\bf R} {\bf S}_{\text{MRS}}^{-1} {\bf R}^\top)^{-1}
 ({\bf r}- {\bf R} {\widehat{\bf \beta}}_{\text{LT1,MRS}}),
\end{align}
where ${\widehat{\bf \beta}}_{\text{LT1,MRS}}$, from \eqref{liu_one} is defined as
${\widehat{\bf \beta}}_{\text{LT1,MRS}} = ({\bf S}_{\text{MRS}} + \I)^{-1} ({\bf S}_{\text{MRS}}+ d \I) {\widehat{\bf \beta}}_{\text{LS,MRS}}$. 
\begin{lemma}\label{moments_srl_rss}
The expected value and covariance of ${\widehat{\bf \beta}}_{\text{SRL,MRS}}$ are given by
\[
\E({\widehat{\bf \beta}}_{\text{SRL,MRS}}) =  {\beta} + {\bf A}_{\text{SRL,MRS}} ({\bf F}_{{\text{MRS}},d} - \I) {\bf S}_{\text{MRS}} {\bf \beta},
\]
\[
\text{Var} ({\widehat{\bf \beta}}_{\text{SRL,MRS}}) = \sigma^2 {\bf A}_{\text{SRL,MRS}} ({\bf F}_{{\text{MRS}},d} 
{\bf S}_{\text{MRS}} {\bf F}_{{\text{MRS}},d}^\top + v^2 {\bf R}^\top {\bf \Omega}^{-1} {\bf R}) {\bf A}_{\text{SRL,MRS}},
\]
where ${\bf A}_{\text{SRL,MRS}}=({\bf S}_{\text{MRS}} + v {\bf R}^\top {\bf \Omega}^{-1} {\bf R})^{-1}$ and
 ${\bf F}_{{\text{MRS}},d} = ({\bf S}_{\text{MRS}} +\I)^{-1} ({\bf S}_{\text{MRS}} +d \I)$.
\end{lemma}
As another method to  deal with collinearity, following \cite{arumairajan2014improvement}, we can use properties of 
ridge estimation based on MRS data in the context of stochastic restricted regression. 
To do so, we replace ${\widehat{\bf \beta}}_{\text{LS,MRS}}$ with 
${\widehat{\bf \beta}}_{\text{R,MRS}}$ in Lemma \ref{mixed_ls_rss}. Hence, the stochastic 
restricted ridge estimator based on MRS data is proposed by
 \begin{align}\label{srr_rss}
{\widehat{\bf \beta}}_{\text{SRR,MRS}} = {\widehat{\bf \beta}}_{\text{R,MRS}} 
+ v {\bf S}_{\text{MRS}}^{-1} {\bf R}^\top ({\bf \Omega} + v {\bf R} {\bf S}_{\text{MRS}}^{-1} {\bf R}^\top)^{-1}
 ({\bf r}- {\bf R} {\widehat{\bf \beta}}_{\text{R,MRS}}),
\end{align}
where ${\widehat{\bf \beta}}_{\text{R,MRS}}$ is defined by \eqref{ridge_rss}.
\begin{lemma}\label{moments_srr_rss}
The expected value and covariance of ${\widehat{\bf \beta}}_{\text{SRR,MSR}}$ are given by
\begin{align*}
\E({\widehat{\bf \beta}}_{\text{SRR,MSR}}) &=  {\beta} + {\bf A}_{\text{SRR,MSR}} ({\bf W}_{\text{MSR}} - \I) {\bf S}_{\text{MSR}} {\bf \beta},\\
\text{Var} ({\widehat{\bf \beta}}_{\text{SRR,MSR}}) &= \sigma^2 {\bf A}_{\text{SRR,MSR}} ({\bf W}_{\text{MSR}} 
{\bf S}_{\text{MSR}} {\bf W}_{\text{MSR}}^\top + v^2 {\bf R}^\top {\bf \Omega}^{-1} {\bf R}) {\bf A}_{\text{SRR,MSR}},
\end{align*}
where ${\bf A}_{\text{SRR,MSR}}= ({\bf S}_{\text{MRS}} + v {\bf R}^\top {\bf \Omega}^{-1} {\bf R})^{-1}$ and  ${\bf W}_{\text{MRS}} = (\I +k {\bf S}_{\text{MRS}}^{-1})^{-1}$.
\end{lemma}
\subsection{Logistic Regression Estimators with RSS}\label{sub:log_rss}
In this section, we investigate the use of MRS data for shrinkage estimators of logistic regression. 
Let ${\bf y}_{\text{MRS}}$ and ${\bf X}_{\text{MRS}}$ 
denote respectively the binary response vector and design matrix with
 $\text{rank}({\bf X}_{\text{MRS}})=p < N$ from MRS data when we ignore the ranking weights, with size $N=nH$, set size $H$ and cycle size $n$.  


As described in Section \ref{sec:srs}, the ML method is the  most common 
approach to estimate of the coefficients of the model.  Similar to Section \ref{sub:log},  
one can apply Newton-Raphson (NR) method \eqref{nr} to MRS data and 
obtain ${\widehat{\bf \beta}}_{\text{ML,MRS}}$.
 Here, we redesign the NR algorithm \eqref{nr} and obtain ${\widehat{\bf \beta}}_{\text{ML,MRS}}$ 
  as a solution to the iteratively re-weighted least square equation
  \begin{align}\label{hosmer_rss}
 {\widehat{\bf \beta}}_{\text{MRS}}^{(l+1)} = \underset{\beta}{\arg\max} ~ ({\bf Z}_{\text{MRS}} 
 - {\bf X}_{\text{MRS}} {\bf\beta})^\top {\bf V}_{\text{MRS}} ({\bf Z}_{\text{MRS}} - {\bf X}_{\text{MRS}} {\bf\beta}),
  \end{align}
  where ${\bf Z}_{\text{MRS}}= \left\{ {\bf X}_{\text{MRS}} {\widehat{\bf \beta}}_{\text{MRS}}^{(l)} 
  + {\bf V}_{\text{MRS}}^{-1} 
  \left[{\bf y}_{\text{MRS}}-{\bar g} ({\bf X}_{\text{MRS}};{\widehat{\bf \beta}}_{\text{MRS}}^{(l)})\right]\right\}$, 
  with ${\bar g} (\cdot)$ 
  represents the vector of the link functions, ${\bf V}_{\text{MRS}}$ is a diagonal matrix with entries 
  \[
  {\bf v}_{i,i} = \exp\left( {\bf x}_{i,\text{MRS}}^\top {\widehat{\bf \beta}}_{\text{MRS}}^{(l)}\right) 
  \left[1+ \exp\left( {\bf x}_{i,\text{MRS}}^\top {\widehat{\bf \beta}}_{\text{MRS}}^{(l)}\right)\right]^{-2},
  \]
  and ${\bf x}_{i,\text{MRS}}$ indicates the $i$th row of design matrix ${\bf X}_{\text{MRS}}, i=1,\ldots, nH$.
  
  While ${\widehat{\bf \beta}}_{\text{ML,MRS}}$ enjoys ranking information from multiple sources, it is still 
  not robust against the collinearity issue. The first solution to the problem is the ridge logistic method \citep{schaefer1984ridge}. 
  Thus, the ridge logistic estimator of ${\bf \beta}$ based on MRS data is defined by
    \begin{align}\label{ridge_log_rss} 
{\widehat{\bf \beta}}_{\text{R,MRS}} 
= ( {\bf X}_{\text{MRS}}^\top {\bf V}_{\text{MRS}} {\bf X}_{\text{MRS}} + k \I)^{-1} 
{\bf X}_{\text{MRS}}^\top {\bf V}_{\text{MRS}} {\bf X}_{\text{MRS}} {\widehat{\bf\beta}_{\text{ML,MRS}}},
\end{align}
where ${\widehat{\bf\beta}_{\text{ML,MRS}}}$ is obtained from \eqref{hosmer_rss}. While small values
 of $k$ are desirable, the ridge logistic estimator may 
 not be able to fully overcome the ill-conditioned matrix $({\bf X}_{\text{MRS}}^\top {\bf V}_{\text{MRS}} 
 {\bf X}_{\text{MRS}})$ \citep{inan2013liu,liu2003using}. Thus, we propose Liu-type logistic estimator 
 with MRS data as follows
    \begin{align}\label{liu_log_rss} 
{\widehat{\bf \beta}}_{\text{LT,MRS}}  =
 ( {\bf X}_{\text{MRS}}^\top {\bf V}_{\text{MRS}} {\bf X}_{\text{MRS}} + k \I)^{-1} ({\bf X}_{\text{MRS}}^\top {\bf V}_{\text{MRS}} {\bf X}_{\text{MRS}} - d \I) {\widehat{\bf\beta}_{\text{ML,MRS}}},
\end{align}
where $k >0$, $d \in \R$. We estimate the shrinkage parameters $k$ and $d$ based on MRS data similar to \cite{inan2013liu}. To this end, we obtin $k=(p+1)/{\widehat{\bf\beta}}_{ML,MRS}^\top {\widehat{\bf\beta}_{ML,MRS}}$ where $p$ 
denotes the number of predictors.
 In addition, we find the optimal $d$ value based on MRS data by minimizing numerically:
 \begin{align} \label{mse_rss} \nonumber
 \text{MSE}({\widehat{\bf \beta}}_{\text{LT,MRS}}) &= 
 tr\left[ 
 ( {\bf X}_{\text{MRS}}^\top {\bf V}_{\text{MRS}} {\bf X}_{\text{MRS}} + k \I) 
 ({\bf X}_{\text{MRS}}^\top {\bf V}_{\text{MRS}} {\bf X}_{\text{MRS}} - d \I) 
 ( {\bf X}_{\text{MRS}}^\top {\bf V}_{\text{MRS}} {\bf X}_{\text{MRS}})^{-1} \right.\\\nonumber
 &~~~~~~ \left. ({\bf X}_{\text{MRS}}^\top {\bf V}_{\text{MRS}} {\bf X}_{\text{MRS}} - d \I) 
 ( {\bf X}_{\text{MRS}}^\top {\bf V}_{\text{MRS}} {\bf X}_{\text{MRS}} + k \I)^{-1}
 \right] \\\nonumber
 &~~+ ||
 ( {\bf X}_{\text{MRS}}^\top {\bf V}_{\text{MRS}} {\bf X}_{\text{MRS}} + k \I)^{-1} 
 ({\bf X}_{\text{MRS}}^\top {\bf V}_{\text{MRS}} {\bf X}_{\text{MRS}} - d \I) 
 ( {\bf X}_{\text{MRS}}^\top {\bf V}_{\text{MRS}} {\bf X}_{\text{MRS}})^{-1}  \\
  &~~~~~~~~ {\bf X}_{\text{MRS}}^\top {\bf V}_{\text{MRS}} {\bar g} ({\bf X}_{\text{MRS}}) -{\bf\beta}||
 \end{align}
 as a function of $d$ for a fixed $k$.

\section{Simulation Studies}\label{sec:num}
  In this section, through three studies, we simulate the performance of the developed methods in 
  estimating the coefficients of linear regression \eqref{reg}, stochastic restricted regression \eqref{reg_sto}
   and logistic regression \eqref{logistic}.  The median 
 RSS results in more efficient estimates in symmetric populations \citep{muttlak1998median}. Because of the
  symmetry of the error term in linear regression models \eqref{reg} and \eqref{reg_sto}, we only focus 
  on shrinkage methods based on median RSS with single observer (MMRS) and median RSS with multiple 
  observers (MMRM) in the first two simulations studies. 

\begin{figure}
\includegraphics[width=1\textwidth,center]{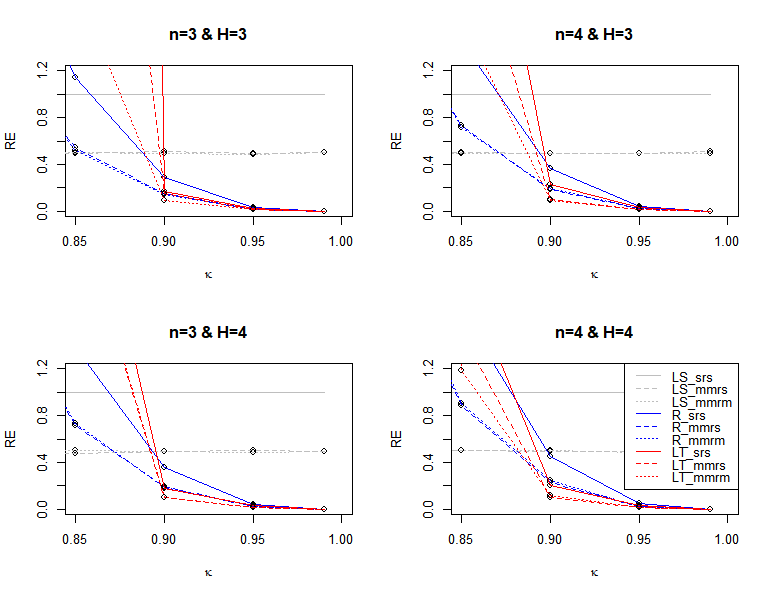}
\caption{\footnotesize{The REs of ${\widehat{\bf\beta}}_{\text{LS}}$, ${\widehat{\bf\beta}}_{\text{R}}$ and ${\widehat{\bf\beta}}_{\text{LT}}$ 
 under SRS, MMRS and MMRM data relative to their ${\widehat{\bf\beta}}_{\text{LS,SRS}}$ counterpart of the same size when $c=1$ with ranking ability $\rho=(1,1,1)$.}}
 \label{sim_reg_all}
\end{figure}
  
   In the first simulation,  we evaluate the  shrinkage estimators  for 
  linear regression in the presence of collinearity. We examine how the proposed RSS estimators 
  are affected by sampling parameters, ranking ability and ties information. 
 To simulate the
   collinearity between predictors, we first generate  $u_{ij}, i=1,\ldots,N;j=1,\ldots,p+1$ from 
   standard normal distribution.  
The predictors are generated by
\begin{align}\label{gen:predictor}
x_{i,j} = (1-\kappa^2)^{1/2} u_{i,j} + \kappa u_{i,p+1} ~~~ i=1,\ldots,N; j=1,\ldots, p, 
\end{align}
 where $\kappa$ accounts for the level of collinearity. 
 Here, we consider the multiple regression with four (i.e., $p=4$) predictors 
 ${\bf x}=({x}_1, {x}_2,{ x}_3,{ x}_4)$  with nominal correlation levels $\kappa=\{0.85,0.9,0.95,0.99\}$.
Using the predictors from \eqref{gen:predictor} and true coefficients ${\bf\beta}_0= (0.25,0.25,0.25,0.25)$,
we generated the vector of responses as $ y_i = {\bf x}_i {\bf\beta}_0 + \epsilon_i$ where $\epsilon_i;i=1,\ldots,N$ 
are generated (independently from predictors) from standard normal distribution. 
There  are various proposals for estimation of ridge and Liu parameters for linear regression. As one of the most 
common approaches, following \cite{hoerl1975ridge}, we obtained the optimal value of $k$ for  ridge estimators 
 by ${\widehat k}_{\text{HKB}} = p {\hat \sigma}/ {\bf\beta}^\top_{LS} {\bf\beta}_{LS}$. From \cite{liu2003using}, 
 we determined the optimal values of $k$ and $d$ for Liu-type estimators as follows:
\begin{align*}
{\widehat k}_{\text{LT}} = \frac{\lambda_1 - 100 \lambda_p}{99}, ~~
{\widehat d}_{\text{LT}} = 
-\frac{ \sum_{j=1}^{p} \lambda_j ( \widehat{\sigma}^2_R - {\widehat k}_{\text{LT}} {\widehat \alpha}^2_{R,j})/ (\lambda_j + {\widehat k}_{\text{LT}})^3}
{ \sum_{j=1}^{p} \lambda_j ( \lambda_, {\widehat \alpha}^2_{R,j} + \widehat{\sigma}^2_R ) / (\lambda_j + {\widehat k}_{\text{LT}})^4}
\end{align*}
where one can obtain ${\widehat \alpha}^2_{R}= (\Lambda +k\I)^{-1} {\bf Z}^\top {\bf y}$ and $\widehat{\sigma}^2_R$ with the estimate of $\sigma^2$ from the ridge method. 
\begin{table}[ht]
\centering
\footnotesize{\begin{tabular}{c|c|cccccc}
  \hline
    $n$ & $\kappa$ &  ${\widehat{\bf \beta}}_{\text{SRR,SRS}}$ &  ${\widehat{\bf \beta}}_{\text{SRR,MMRS}}$
    &  ${\widehat{\bf \beta}}_{\text{SRR,MMRM}}$ &  ${\widehat{\bf \beta}}_{\text{SLR,SRS}}$
    &  ${\widehat{\bf \beta}}_{\text{SLR,MMRS}}$ &  ${\widehat{\bf \beta}}_{\text{SLR,MMRM}}$ \\ 
  \hline
   $3$& 0.75 & 0.528 & 0.363 & 0.368 & 1.231 & 0.703 & 0.708 \\ [-1ex]
       & 0.80 & 0.365 & 0.283 & 0.279 & 1.214 & 0.746 & 0.669 \\ [-1ex]
       & 0.85 & 0.292 & 0.255 & 0.256 & 1.193 & 0.787 & 0.703 \\ [-1ex]
       & 0.90 & 0.259 & 0.249 & 0.249 & 3.229 & 2.769 & 4.751 \\ [-1ex]
       & 0.95 & 0.256 & 0.252 & 0.252 & 266 & 39.156 & 49.129 \\ [-1ex]
       & 0.99 & 0.261 & 0.257 & 0.257 & 0.529 & 0.949 & 0.302 \\ 
   \cline{1-8}
   $4$& 0.75 & 0.502 & 0.339 & 0.336 & 1.311 & 0.711 & 0.714 \\ [-1ex]
    &0.80 & 0.337 & 0.261 & 0.262 & 1.281 & 0.683 & 0.685 \\ [-1ex]
    &0.85 & 0.278 & 0.248 & 0.247 & 1.239 & 0.648 & 0.647 \\ [-1ex]
    &0.90 & 0.255 & 0.246 & 0.246 & 1.103 & 0.625 & 0.586 \\ [-1ex]
    &0.95 & 0.255 & 0.251 & 0.251 & 9477 & 6.208 & 56.030 \\ [-1ex]
    &0.99 & 0.259 & 0.256 & 0.256 & 0.712 & 0.330 & 0.361 \\ 
  \hline
\end{tabular}
\caption{\footnotesize{The REs of ${\widehat{\bf\beta}}_{\text{SRR}}$, ${\widehat{\bf\beta}}_{\text{SLR}}$ under SRS, MMRS and MMRM data relative to their
${\widehat{\bf\beta}}_{\text{LS,SRS}}$ counterpart of the same size when $H =3$ and $c=1$ with ranking ability $\rho=(1,1,1)$.}}
\label{tab_sim_sto}}
\end{table}

To study the impact of ranking ability, from \cite{dell1972ranked}, we generated external observer $R_l (l=1,\ldots,K)$ by 
\[
R_{i,l}= (1-\rho^2_l)^{1/2} y_i + \rho u_i, ~~ i=1,\ldots,N; l=1,\ldots,K,
\]
where $\rho_l=\text{Cor}(R_l,{\bf y})$ and $u_i$ comes from an independent standard normal distribution. 
 In this simulation, we considered three ranking levels $\rho=\{0.5,0.9,1\}$ when we collect MMRS data. 
 In addition, we considered six ranking combinations of $\rho$ to construct MMRM data with two and three  observers. 
 The DPS model with tie parameter $c=\{0.5,1,2\}$ were also applied to record the ties information between units 
 in MMRS and MMRM data. 
 Accordingly, we generated  $\mathbb{C}=10000$ predictors and responses under SRS, MMRS and MMRM data of the 
 same sizes $N=\{9,12\}$ with set size $H=\{3,4\}$ and calculated the MSE of the estimates as
 $
 \text{MSE}({\widehat{\bf\beta}}) = 1/\mathbb{C} \sum_{l=1}^{\mathbb{C}} ({\widehat{\bf\beta}} -{\bf\beta}_0)^\top ({\widehat{\bf\beta}} -{\bf\beta}_0).
 $
 We compute the  efficiency (RE) of shrinkage estimator ${\widehat{\bf \beta}}$ relative
to ${\widehat{\bf \beta}}_{\text{LS,SRS}}$ to measure the performance of ${\widehat{\bf \beta}}$. 
The $\text{RE}({\widehat{\bf \beta}})$ is given by 
\[
\text{RE}({\widehat{\bf \beta}}) = \frac{\text{MSE}({\widehat{\bf \beta}})}{\text{MSE}({\widehat{\bf \beta}}_{\text{LS,SRS}})},
\]
where $\text{RE} < 1$ indicates the superiority of ${\widehat{\bf \beta}}$ in the estimation of 
coefficients of regression in the presence of collinearity.

\begin{table}[ht]
\centering
\footnotesize{\begin{tabular}{c|c|ccc|ccc}
  \hline
    &  & &$n=4, H=6$ &  & &$n=2, H=12$ &  \\
  \cline{3-8}
  $\phi$& Estimator  & Median & 2.5\% & 97.5\% & Median & 2.5\% & 97.5\% \\ 
  \hline
0.95& ${\widehat{\bf\beta}}_{\text{LS,SRS}}$ & 2657.26 & 224.40 & 26303 & 2511.01 & 229.39 & 25043 \\ [-1ex]
    & ${\widehat{\bf\beta}}_{\text{R,SRS}}$  & 307.39  & 36.46 & 3900 & 285.41 & 35.66 & 3353 \\ [-1ex]
    & ${\widehat{\bf\beta}}_{\text{LT,SRS}}$ & 58.02   & 11.99 & 687 & 54.77 & 11.46 & 627 \\ 
    & ${\widehat{\bf\beta}}_{\text{LS,RSS}}$ & 2541.31 & 266.43 & 20646 & 2455.98 & 242.12 & 19710 \\ [-1ex]
    & ${\widehat{\bf\beta}}_{\text{R,RSS}}$  & 287.91  & 39.35 & 3624 & 283.34 & 37.71 & 3252 \\ [-1ex]
    & ${\widehat{\bf\beta}}_{\text{LT,RSS}}$ & 56.51   & 12.53 & 701 & 56.60 & 11.99 & 692 \\ 
    & ${\widehat{\bf\beta}}_{\text{LS,MRS}}$ & 2390.56 & 251.69 & 20043 & 2266.78 & 243.14 & 19860\\[-1ex]
    & ${\widehat{\bf\beta}}_{\text{R,MRS}}$  & 288.24  & 38.74 & 3162 & 260.80 & 36.59 & 2782 \\ [-1ex]
    & ${\widehat{\bf\beta}}_{\text{LT,MRS}}$ & 56.88   & 12.47 & 637 & 51.22 & 11.74 & 597 \\ 
    & ${\widehat{\bf\beta}}_{\text{LS,MMR}}$ & 4334.09 & 455.25 & 40774 & 4065.53 & 320.09 &566201 \\[-1ex]
    & ${\widehat{\bf\beta}}_{\text{R,MMR}}$  & 477.85 & 57.98 & 4399 & 439.38 & 48.57 & 4805 \\ [-1ex]
    & ${\widehat{\bf\beta}}_{\text{LT,MMR}}$ & 94.16 & 15.78 & 874 & 84.68 & 14.05 & 885 \\ 
    \hline
0.98& ${\widehat{\bf\beta}}_{\text{LS,SRS}}$ & 4399.16 & 320.80 & 50018 & 4363.90 & 307.59 & 53791 \\ [-1ex]
    & ${\widehat{\bf\beta}}_{\text{R,SRS}}$  & 308.64  & 40.60 & 3509 & 307.09 & 39.57 & 3366 \\ [-1ex]
    & ${\widehat{\bf\beta}}_{\text{LT,SRS}}$ & 54.59   & 12.58 & 663 & 55.18 & 12.55 & 649 \\ 
    & ${\widehat{\bf\beta}}_{\text{LS,RSS}}$ & 4216.01 & 291.41 & 43561 & 4140.31 & 332.66 & 44098 \\ [-1ex]
    & ${\widehat{\bf\beta}}_{\text{R,RSS}}$  & 295.37  & 40.90 & 3084 & 299.80 & 40.43 & 3016 \\ [-1ex]
    & ${\widehat{\bf\beta}}_{\text{LT,RSS}}$ & 53.32   & 13.29 & 596 & 54.13 & 12.95 & 577 \\     
    & ${\widehat{\bf\beta}}_{\text{LS,MRS}}$ & 3941.06 & 308.70 & 42562 & 3909.15 & 284.66 & 43700 \\ [-1ex]
    & ${\widehat{\bf\beta}}_{\text{R,MRS}}$  & 271.68  & 40.51 & 2818 & 293.45 & 36.41 & 2996 \\ [-1ex]
    & ${\widehat{\bf\beta}}_{\text{LT,MRS}}$ & 50.83   & 12.60 & 564 & 52.28 & 11.96 & 560 \\ 
    & ${\widehat{\bf\beta}}_{\text{LS,MMR}}$ & 7352.59 & 551.27 & 84925 & 6771.71 & 438 & 102627 \\[-1ex]
    & ${\widehat{\bf\beta}}_{\text{R,MMR}}$  & 499.72  & 60.25 & 4648 & 457.79 & 54.14 & 5065 \\ [-1ex]
    & ${\widehat{\bf\beta}}_{\text{LT,MMR}}$ & 86.96 & 16.64 & 923 & 79.29 & 15.32 & 902 \\ 
   \hline
\end{tabular}
\caption{\footnotesize{The median and 95\% CI for the SSE of estimators of coefficients of the logistic model 
under SRS, RSS, MRS and MMR data of the same size when $\eta=0.95$, $c=0.2$ and ranking ability $(\rho_1,\rho_2)=(0.95,0.95)$.}}
\label{tab_sim_log_r95_good}}
\end{table}

\begin{figure}
\includegraphics[width=1\textwidth,center]{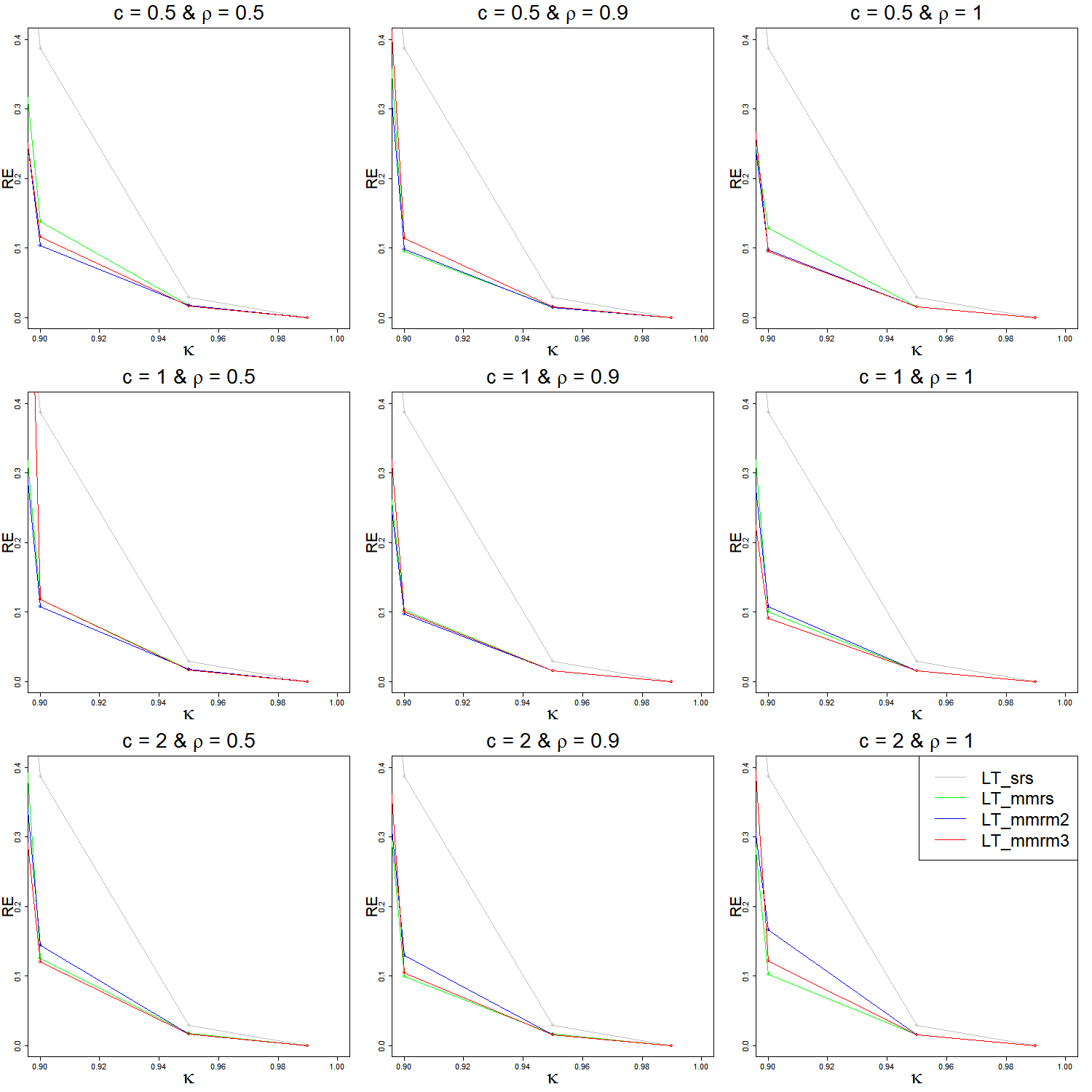}
\caption{\footnotesize{The REs of ${\widehat{\bf\beta}}_{\text{LT}}$ under SRS, RSS, MRS and MMR data 
relative to their ${\widehat{\bf\beta}}_{\text{LS,SRS}}$ counterpart of the same size when $H=3$ and $n=4$.}}
 \label{sim_reg_LT_H3}
\end{figure}

\begin{figure}
\includegraphics[width=1\textwidth,center]{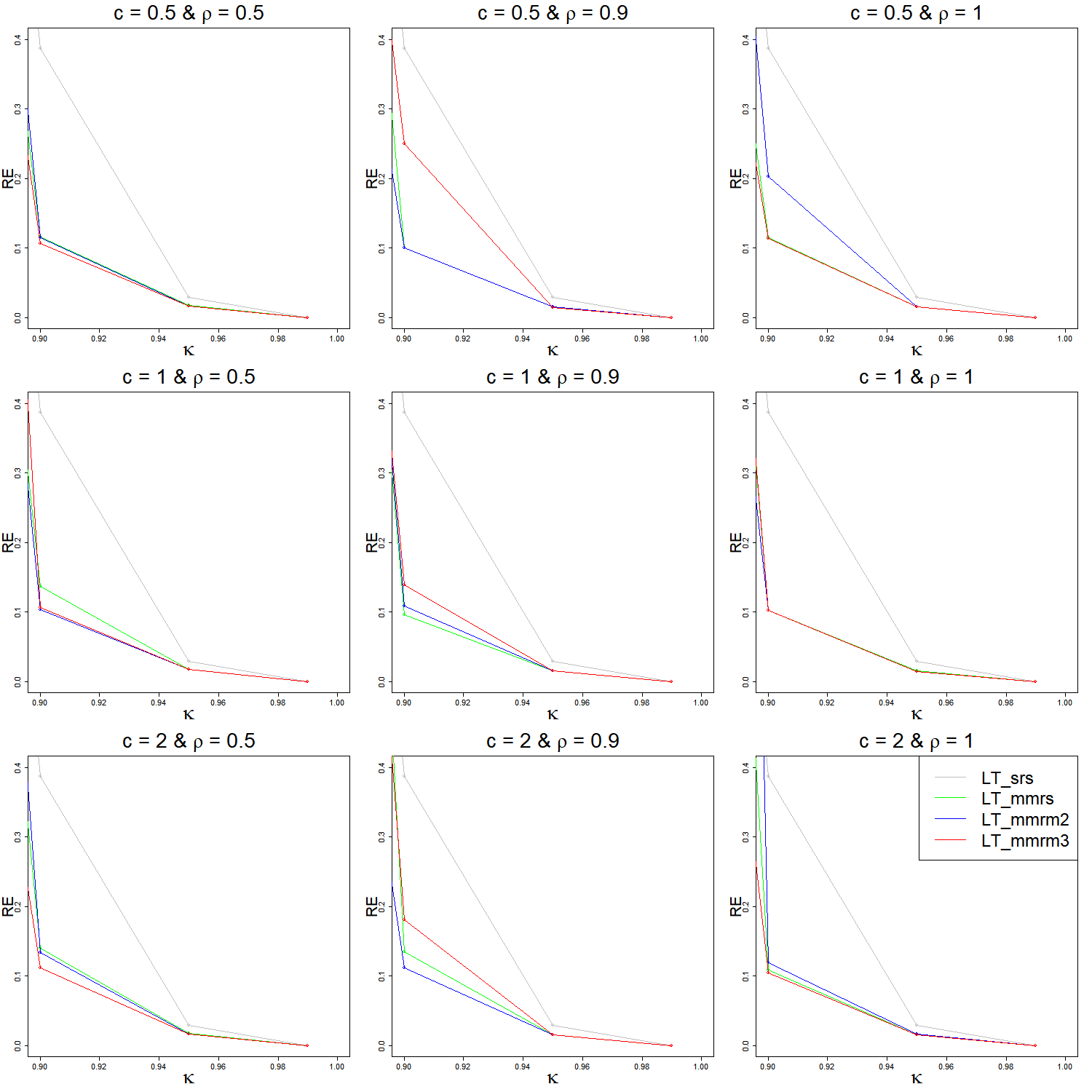}
\caption{\footnotesize{The REs of ${\widehat{\bf\beta}}_{\text{LT}}$ under SRS, RSS, MRS and MMR data relative to their ${\widehat{\bf\beta}}_{\text{LS,SRS}}$ counterpart of the same size when $H=4$ and $n=3$.}}
 \label{sim_reg_LT_H4}
\end{figure}

\begin{figure}
\includegraphics[width=1\textwidth,center]{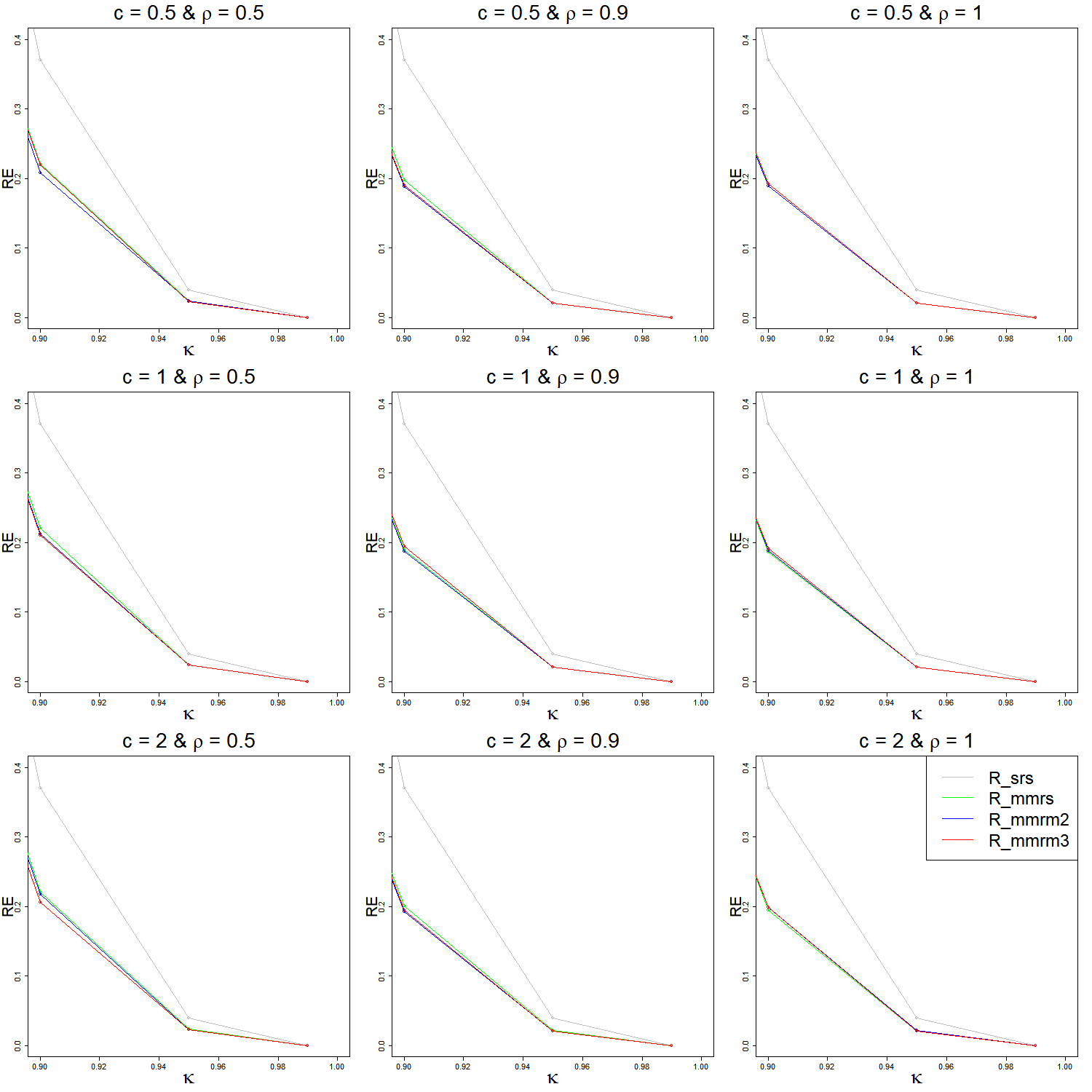}
\caption{\footnotesize{The REs of ${\widehat{\bf\beta}}_{\text{R}}$ under SRS, RSS, MRS and MMR data relative to their ${\widehat{\bf\beta}}_{\text{LS,SRS}}$ counterpart of the same size when $H=3$ and $n=4$.}}
 \label{sim_reg_R_H3}
\end{figure}

\begin{figure}
\includegraphics[width=1\textwidth,center]{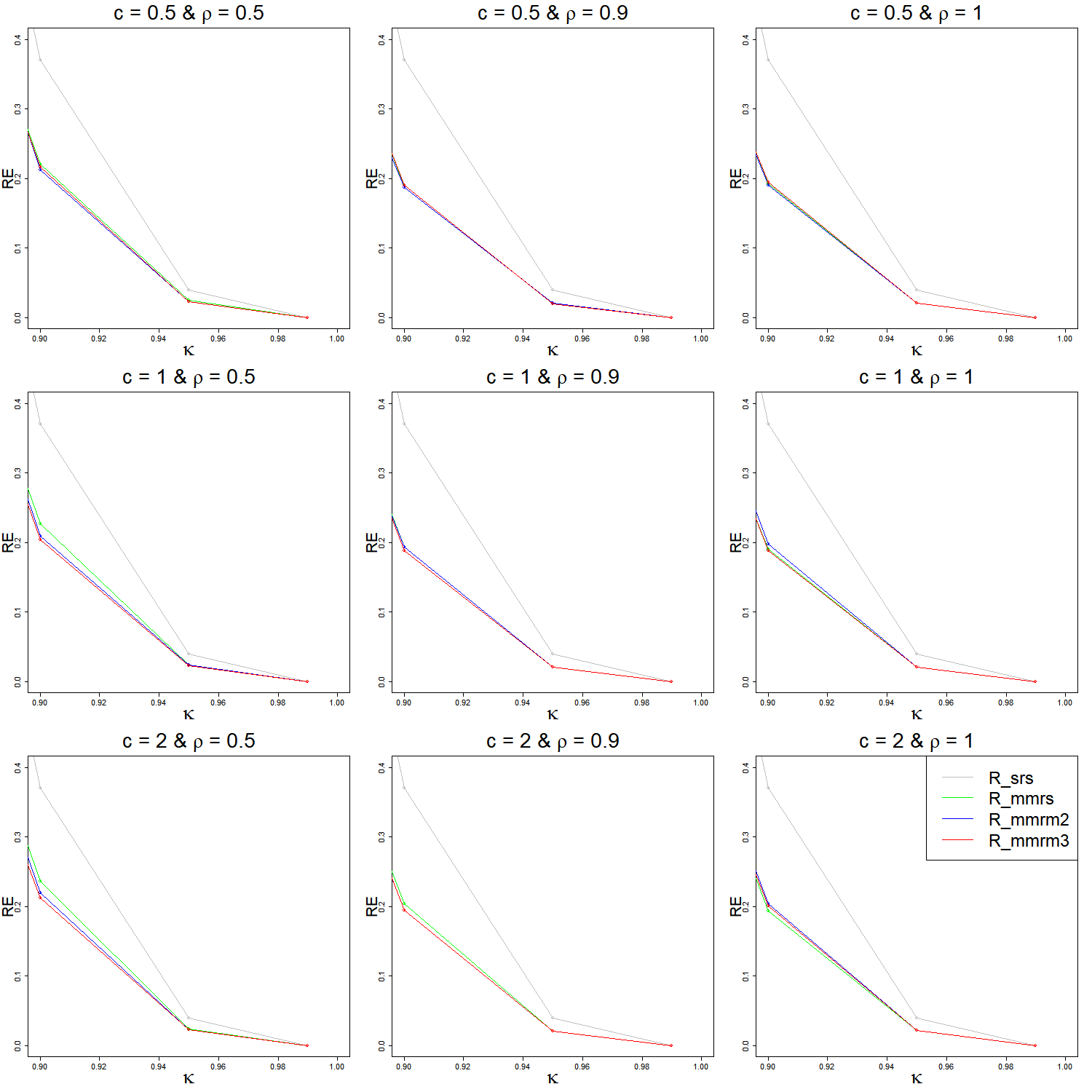}
\caption{\footnotesize{The REs of ${\widehat{\bf\beta}}_{\text{R}}$ under SRS, RSS, MRS and MMR data relative to their ${\widehat{\bf\beta}}_{\text{LS,SRS}}$ counterpart of the same size when $H=4$ and $n=3$.}}
 \label{sim_reg_R_H4}
\end{figure}

Figures \ref{sim_reg_all} --\ref{sim_reg_R_H4} show the results of this simulation study. 
It is evident that the biased shrinkage estimators ${\widehat{\bf \beta}}_{\text{R}}$ and 
${\widehat{\bf \beta}}_{\text{LT}}$ outperform  unbiased estimator ${\widehat{\bf \beta}}_{\text{LS}}$ 
when the regression model suffers from high collinearity problem.
The Liu estimator ${\widehat{\bf \beta}}_{\text{LT,MMRM}}$ shows the best performance 
when collinearity is sever $(\kappa \ge 0.9)$ while the the ridge estimator ${\widehat{\bf \beta}}_{\text{R,MMRM}}$ 
 is recommended when $\kappa \in [0.85,.9)$.  
Figures \ref{sim_reg_LT_H3} -- \ref{sim_reg_R_H4} demonstrate the performance of shrinkage estimators under
 MMR data with 1, 2 and 3 observers. 
The efficiency of  Liu and Ridge estimators with MMRM data (relative to their counterparts 
based on MMRS data) improve as the ranking ability $\rho$ increases from $0.5$ to $1$. The RE of 
multi-observer estimators improve further as the tie parameter $c$ increases from 0.5 to 1; however 
the RE decreases as $c$ increases from 1 to 2 where many tied ranks are produced in the MMRM data collection. 
 
 In the second simulation study, we investigated the performance of SRS, MMRM and MMRS shrinkage 
 estimators for stochastic restricted regression \eqref{reg_sto}. 
 Similar to \cite{arumairajan2014improvement}, we chose the ${\bf\beta}_0=(0.6455,0.0896,0.1436,0.1526)$ as the true coefficients of regression model.  
 To implement \eqref{reg_sto}, Similar to \cite{arumairajan2014improvement}, we  simulated one restriction with ${\bf R}=(1,-2,-2,-2)$ and  
  ${r}=0$, $v=1$ and generated error term ${\bf e}$  from a normal distribution $N(0,0.0015)$.  
 In addition, we generated the SRS and MMRM 
 and MMRS data and computed the RE of shrinkage estimators \eqref{srl_rss} and \eqref{srr_rss} in a similar 
 manner as first simulation study. 
 Tables \ref{tab_sim_sto}, \ref{tab_sim_sto_H4} and \ref{tab_sim_sto_c_r} show the result of the RE of 
 estimators \eqref{srl_rss} and \eqref{srr_rss} based on MMRS and MMRM data when 
 $\kappa \in \{ .75,.80,.85,.90,0.95,.99\}$ and tie-parameter $c\in \{.1,.5,1\}$.
 It is apparent that almost all RSS-based shrinkage estimators outperform their SRS counterparts. 
 Comparing RSS estimators, one observes that the multi-observer methods more efficiently estimate the coefficients of restricted regression in the presence of collinearity than their singe-observer competitors. 
 As reported by \cite{arumairajan2014improvement}, 
 ${\widehat{\bf \beta}}_{\text{SLR}}$ becomes more unstable when collinearity is severe in the model. Therefore,  
 if one has access to multiple decent observers, ${\widehat{\bf \beta}}_{\text{SRR,MMRM}}$ is always recommended to deal with collinearity in restricted regression; otherwise, we recommend  ${\widehat{\bf \beta}}_{\text{SLR,MMRS}}$. 
Interestingly, Table \ref{tab_sim_sto_H4} indicates that  ${\widehat{\bf \beta}}_{\text{SLR,MMRM}}$ gets more reliable (in the presence of sever collinearity) when more tied ranks are declared in MMRM data collection. 

 \begin{table}
\centering
\footnotesize{\begin{tabular}{c|ccc|ccc}
  \hline
 &  & $n=4,H=6$ & &  & $n=2,H=12$ &  \\
  \cline{2-7}
 & Median & 2.5\% & 97.5\% & Median & 2.5\% & 97.5\% \\ 
  \hline
  ${\widehat{\bf\beta}}_{\text{LS,SRS}}$& 0.621 & 0.273 & 0.969 & 0.612 & 0.265 & 0.959 \\ [-1ex]
  ${\widehat{\bf\beta}}_{\text{R,SRS}}$ & 0.136 & 0.070 & 0.201 & 0.136 & 0.070 & 0.202 \\ [-1ex]
  ${\widehat{\bf\beta}}_{\text{LT,SRS}}$& 0.133 & 0.071 & 0.195 & 0.133 & 0.070 & 0.196 \\ 
  ${\widehat{\bf\beta}}_{\text{R,RSS}}$ & 0.136 & 0.070 & 0.201 & 0.136 & 0.070 & 0.202 \\ [-1ex]
  ${\widehat{\bf\beta}}_{\text{LT,RSS}}$ & 0.133 & 0.070 & 0.195 & 0.133 & 0.071 & 0.196 \\ 
  ${\widehat{\bf\beta}}_{\text{R,MRS}}$ & 0.136 & 0.070 & 0.201 & 0.136 & 0.070 & 0.201 \\ [-1ex]
  ${\widehat{\bf\beta}}_{\text{LT,MRS}}$ & 0.133 & 0.070 & 0.195 & 0.133 & 0.071 & 0.195 \\ 
  ${\widehat{\bf\beta}}_{\text{R,MMR}}$ & 0.131 & 0.070 & 0.193 & 0.129 & 0.070 & 0.187 \\ [-1ex]
  ${\widehat{\bf\beta}}_{\text{LT,MMR}}$ & 0.128 & 0.070 & 0.185 & 0.124 & 0.070 & 0.179 \\ 
   \hline
\end{tabular}
\caption{\footnotesize{The Median and 95\% CI for the SSE of methods in estimating the bone mineral 
population via regression model \eqref{reg} 
under SRS, RSS, MRS and MMR data of the same size when $c=0.1$.}}
\label{tab_real_reg_c}}
\end{table}

In the third simulation study, we examine the performance of  RSS-based shrinkage methods in estimating the 
coefficients of logistic regression \eqref{logistic}. Similar to \cite{inan2013liu}, we simulated the logistic 
regression with $p=4$ predictors where collinearity is incorporated into the model by two parameter $\phi$ and $\eta$. 
The four predictors are generated by
\[
x_{i,j_1} = (1-\phi^2)^{1/2} u_{i,j_1} + \phi u_{i,p+1}, ~~~
x_{i,j_2} = (1-\eta^2)^{1/2} u_{i,j_2} + \eta u_{i,p+1}, 
\]
where $j_1=1,2$ and $j_2=3,4$ and errors ${u}_{i,j}$ are generated independently from standard normal distribution. 
In this simulation, we set the true coefficients of logistic regression as ${\bf\beta}_0= (-0.2,1.3,0.8,-0.3,-0.9)$ with 
$\phi=\{0.95,0.98\}$ and $\eta=\{0.95,0.98\}$. To generate external observers in RSS data collection, we followed 
the data generation of \cite{zamanzade2018proportion}. To do so, for a given binary response $y$ and predictors 
${\bf x}$, the external observer $R$ is generated  as $R|y=0\sim N(0,1)$ and 
\[
R|y=1 \sim N \left(\rho/\sqrt{(1-\rho^2) g({\bf x},{\bf\beta}_0) (1-g({\bf x},{\bf\beta}_0))},1 \right),
\]
where $\rho=\text{Cor}(R,y)$  and $g(\cdot)$ is given by \eqref{logistic}. We considered $\rho=\{0.75,0.95\}$ to 
obtain the ranking information in RSS (with single observer),  MRS and MMR samples. We also set tie-parameter 
${c}={0.2}$ in DPS model to explore the impact  of the ties information on the proposed shrinkage estimators. 
 We used the measure $\text{SSE} ({\widehat{\bf\beta}})=({\widehat{\bf\beta}}-{\bf\beta}_0)^\top ({\widehat{\bf\beta}}-{\bf\beta}_0)$ 
 to evaluate the performance of coefficient estimate ${\widehat{\bf\beta}}$ in logistic regression with collinearity problem. 
 Accordingly, the shrinkage estimation procedures were replicated 10000 times under SRS, RSS, MRS and MMR data of size $N=24$ with $H=\{6,12\}$. 
 We then computed the median and 95\% non-parametric confidence interval (CI) for the SSE of the estimates. The lower and upper bands of each CI were calculated by 2.5 and 97.5 percentiles  of the SSEs, respectively.

\begin{table}[ht]
\centering
\footnotesize{\begin{tabular}{c|ccc|ccc}
\hline
 &  & $n=4,H=6$ &  &  & $n=2,H=12$ &  \\
  \cline{2-7}
 & Median & 2.5\% & 97.5\% & Median & 2.5\% & 97.5\% \\
  \hline
  ${\widehat{\bf\beta}}_{\text{LS,SRS}}$ & 42.47 & 0.22 & 982.01 & 43.08 & 0.26 & 821.81 \\ [-1ex]
  ${\widehat{\bf\beta}}_{\text{R,SRS}}$ & 2.94 & 0.17 & 384.88 & 3.00 & 0.16 & 296.15 \\ [-1ex]
  ${\widehat{\bf\beta}}_{\text{LT,SRS}}$ & 2.38 & 0.27 & 40.77 & 2.38 & 0.27 & 37.82 \\ 
  ${\widehat{\bf\beta}}_{\text{R,RSS}}$ & 2.88 & 0.12 & 322.69 & 2.75 & 0.11 & 305.16 \\ [-1ex]
  ${\widehat{\bf\beta}}_{\text{LT,RSS}}$ & 2.38 & 0.14 & 38.09 & 2.38 & 0.15 & 33.74 \\ 
  ${\widehat{\bf\beta}}_{\text{R,MRS}}$ & 2.88 & 0.11 & 318.75 & 2.70 & 0.10 &  299.59\\ [-1ex]
  ${\widehat{\bf\beta}}_{\text{LT,MRS}}$ & 2.38 & 0.18 & 37.22 & 2.38 & 0.16 & 36.67 \\ 
   \hline
\end{tabular}
\caption{\footnotesize{The Median and 95\% CI for the SSE of methods in the estimation of the bone mineral 
population via logistic model \eqref{logistic} 
under SRS, RSS and MRS data of the same size when $c=0.1$.}}
\label{tab_real_log_c}}
\end{table}

 The results of this simulation study are shown in Tables 
 \ref{tab_sim_log_r95_good}, \ref{tab_sim_log_r98_good} - \ref{tab_sim_log_r98_bad}.
 Similar to previous simulation studies, the LS estimates  are dramatically influenced by collinearity 
 in the logistic regression; unlike LS estimators, there is a significant reduction in the SSE of Liu and ridge estimators.
 When the collinearity is high, the Liu estimators outperform their ridge competitors in estimating the 
  logistic regression coefficients. 
 Shrinkage estimators based on RSS and MRS data almost always outperform their SRS counterparts. Unlike 
 linear regression where estimators with MMR data were preferred,   the RSS and MRS data results in more 
 reliable shrinkage estimates  of coefficients of logistic regression. Hence, we only presented the results 
 based on RSS and MRS data for analysis of logistic regression. While the median SSE of 
 ${\widehat{\bf\beta}}_{\text{LT,MRS}}$ and ${\widehat{\bf\beta}}_{\text{LT,RSS}}$ are close, the multi-observer 
  ${\widehat{\bf\beta}}_{\text{LT,MRS}}$ results in the shortest CIs of the SSEs. Interestingly,
   it is observed that the efficiency of ${\widehat{\bf\beta}}_{\text{LT,MRS}}$ and ${\widehat{\bf\beta}}_{\text{R,MRS}}$ 
    grows further as the collinearity gets more severe  in the logistic regression. 
 When the ranking ability is strong,   the performance of shrinkage estimators based on RSS and MRS data
  is improved further as the set size increases.  
  In short, when one has access to multiple observers with decent ranking abilities,
   ${\widehat{\bf\beta}}_{\text{LT,MRS}}$ is recommended to estimate the coefficients of logistic regression in the presence of collinearity. 
 
\section{Bone Mineral Data Analysis}\label{sec:real}
 As a bone metabolic disorder problem, osteoporosis happens when the density of the patient's 
 bone structure decreases considerably. 
 The disease occurs without a major symptom; that is why it is often called a silent thief.
  There are various characteristics such as sex, age and BMI associated with osteoporosis 
 \citep{felson1993effects,seeman1983risk}. 
 Based on the Korean NHANES survey, around 35\% of women aged 50 
 and older in South Korea suffer from osteoporosis while this proportion becomes less than 8\% 
 for men in that age group \citep{lim2016comparison}. In the case of age, it is 
 known that the bone mineral density increases until age 30s and then decreases as the individual ages \citep{black1992axial}. 
  
  Bone mineral density (BMD) is one of the most reliable factors in determining bone disorder. BMDs, given as T-scores, 
  are compared with a BMD norm to determine the bone disorder status of an individual. If the BMD falls lower 
  than -2.5 SD  from the norm, the status is diagnosed as osteoporosis.  Although measuring BMD is costly and 
  time-consuming, practitioners have access to several easy-to-measure characteristics about the patients, such as 
  demographic information and BMD scores from previous surveys. We believe practitioners can use a multi-observer 
  RSS scheme, translate these characteristics into ranking information to more efficiently  estimate the bone disorder population. 
  Here, we apply RSS, MRS, and MMR samples to analyze bone mineral data obtained from the 
  National Health and Nutritional Examination Survey (NHANES III) conducted with over 33999 American adults by CDC 
  between 1988-1994. The survey includes 241 white women aged 50 and older who participated in two bone 
  examinations. We treat these female participants as our population. Here, we apply the developed shrinkage 
  methods to analyze the BMD data in two numerical studies in the context of linear regression and logistic regression.
  
  In the first study, we considered total BMD (TOBMD) scores from the second bone examination as the response 
  variable of the linear regression.  Weight and BMI characteristics were treated as the two predictors of the model
   where correlation level ${\kappa=.91}$ 
 indicates the collinearity issue in the regression. We treated the TOBMD and INBMD measurements of the first bone 
 examination as two observers with $(\rho_1,\rho_2)=(0.97,0.95)$ for ranking purposes. We replicated shrinkage estimates 10000 times 
  under SRS, RSS, MRS and MMR data of size $N=24$ with $H=\{6,12\}$ and $c=0.1$ and  computed the 
 median and 95\%  CI for SSE as described in Section \ref{sec:num}.
 Table \ref{tab_real_reg_c} illustrates the results of this analysis. 
 Due to the symmetry in linear regression, we see ${\widehat{\bf\beta}}_{\text{LT,MMR}}$ and ${\widehat{\bf\beta}}_{\text{R,MMR}}$
  estimate the true coefficients of the bone mineral linear regression more efficiently; hence they are recommend in this analysis.
   The performance of MMR estimators also improves as the set size increases from 6 to 12 while the sample size remains the same. 
  
  Although the population in the second analysis is the same as in the first, we translated the TOMBD scores 
  from the second examination into binary osteoporosis status. We compared the TOBMD of each individual with the 
  TOBMD norm obtained from individuals aged 20-30. If the BMD was larger than -2.5 SD of the norm, we assign $y=1$ 
  (i.e., normal status); otherwise 
 $y=0$ (i.e., osteoporosis status). The TRBMD and FNBMD measurements from the first bone examination with $\kappa=0.80$ 
 were also considered the two predictors of the logistic regression. We ranked the patients with TOBMD and INBMD measurements
  of the first bone examination as two observers with $(\rho_1,\rho_2)=(0.97,0.96)$. 
 Similar to the first analysis, we computed the median and 95\% CI for SSE of shrinkage estimators based on SRS, RSS and MRS
 samples of size $N=24$ with $H=\{6,12\}$ and $c=0.1$.
 The results are shown in Table \ref{tab_real_log_c}. The shrinkage methods result in a considerable reduction in the 
 length of CIs for SSEs in the presence of collinearity. We also see that the RSS and MRS shrinkage estimators almost 
 always outperform the SRS estimators. Similar to the first analysis,  when practitioners have access to decent observers,
 the performance of RSS-based shrinkage estimators can be improved as the set size increases. 
  


\section{Summary and Concluding Remarks}\label{sec:sum}
In many applications such as osteoporosis research, measuring the  variable of interest is obtained through
 an expensive and time-consuming process; however, practitioners have access to many inexpensive and 
 easy-to-measure characteristics about the individuals. In these situations, ranked set sampling, as an 
 alternative to commonly-used simple random sampling, can translate these characteristics into data 
 collection, providing more representative samples from the population. Collinearity is a common challenge
  in linear models that leads to unreliable coefficient estimates under the least square method and 
  consequently misleading information.  \cite{ebegil2021some} recently proposed ridge and Liu-type estimates
   under RSS data to handle the collinearity; however, the developed RSS estimators can not enjoy multiple
    ranking sources and ties information in data collection. 
Despite the importance of logistic regression in medical studies, no research has investigated the RSS shrinkage
 estimators to deal with the collinearity in logistic regression and stochastic restricted regression.  
 In this manuscript, we developed the Liu-type and ridge estimation methods under multi-observer RSS data to
  estimate the coefficients of the linear regression, stochastic restricted regression and logistic 
  regression in the presence of collinearity. When the collinearity level is high and practitioners have access
   to multiple decent observers, the ${\widehat{\bf \beta}}_{\text{LT,MMRM}}$ 
   is recommended to handle the estimation problem in linear regression and stochastic restricted regression models. In the case of
     high collinearity in the logistic regression model,  ${\widehat{\bf \beta}}_{\text{LT,MRS}}$ shows the most 
     reliable coefficient estimates. 
     
\section*{Acknowledgment}  
Armin Hatefi acknowledges the research support of the Natural Sciences and Engineering Research Council of Canada (NSERC).
\nocite{*}
\bibliographystyle{plainnat}
{\small\bibliography{SMR-bib}

\begin{thebibliography}{40}
\providecommand{\natexlab}[1]{#1}
\providecommand{\url}[1]{\texttt{#1}}
\expandafter\ifx\csname urlstyle\endcsname\relax
  \providecommand{\doi}[1]{doi: #1}\else
  \providecommand{\doi}{doi: \begingroup \urlstyle{rm}\Url}\fi

\bibitem[Alvandi and Hatefi(2021)]{alvandi2021estimation}
Amirhossein Alvandi and Armin Hatefi.
\newblock Estimation of ordinal population with multi-observer ranked set
  samples using ties information.
\newblock \emph{Statistical Methods in Medical Research}, 2021.

\bibitem[Amiri et~al.(2014)Amiri, Jozani, and Modarres]{amiri2014resampling}
Saeid Amiri, Mohammad~Jafari Jozani, and Reza Modarres.
\newblock Resampling unbalanced ranked set samples with applications in testing
  hypothesis about the population mean.
\newblock \emph{Journal of agricultural, biological, and environmental
  statistics}, 19\penalty0 (1):\penalty0 1--17, 2014.

\bibitem[Arumairajan et~al.(2014)Arumairajan, Wijekoon,
  et~al.]{arumairajan2014improvement}
Sivarajah Arumairajan, Pushpakanthie Wijekoon, et~al.
\newblock Improvement of ridge estimator when stochastic restrictions are
  available in the linear regression model.
\newblock \emph{Journal of Statistical and Econometric Methods}, 3\penalty0
  (1):\penalty0 35--48, 2014.

\bibitem[Black et~al.(1992)Black, Cummings, Genant, Nevitt, Palermo, and
  Browner]{black1992axial}
Dennis~M Black, Steven~R Cummings, Harry~K Genant, Michael~C Nevitt, Lisa
  Palermo, and Warren Browner.
\newblock Axial and appendicular bone density predict fractures in older women.
\newblock \emph{Journal of Bone and Mineral Research}, 7\penalty0 (6):\penalty0
  633--638, 1992.

\bibitem[Bliuc et~al.(2009)Bliuc, Nguyen, Milch, Nguyen, and
  Eisman]{bliuc2009mortality}
Dana Bliuc, Nguyen~D Nguyen, Vivienne~E Milch, Tuan~V Nguyen, and John~A
  Eisman.
\newblock Mortality risk associated with low-trauma osteoporotic fracture and
  subsequent fracture in men and women.
\newblock \emph{Jama}, 301\penalty0 (5):\penalty0 513--521, 2009.

\bibitem[Chen et~al.(2005)Chen, Stasny, and Wolfe]{chen2005ranked}
Haiying Chen, Elizabeth~A Stasny, and Douglas~A Wolfe.
\newblock Ranked set sampling for efficient estimation of a population
  proportion.
\newblock \emph{Statistics in medicine}, 24\penalty0 (21):\penalty0 3319--3329,
  2005.

\bibitem[Cummings et~al.(1995)Cummings, Nevitt, Browner, Stone, Fox, Ensrud,
  Cauley, Black, and Vogt]{cummings1995risk}
Steven~R Cummings, Michael~C Nevitt, Warren~S Browner, Katie Stone, Kathleen~M
  Fox, Kristine~E Ensrud, Jane Cauley, Dennis Black, and Thomas~M Vogt.
\newblock Risk factors for hip fracture in white women.
\newblock \emph{New England journal of medicine}, 332\penalty0 (12):\penalty0
  767--774, 1995.

\bibitem[Dell and Clutter(1972)]{dell1972ranked}
TR~Dell and JL~Clutter.
\newblock Ranked set sampling theory with order statistics background.
\newblock \emph{Biometrics}, pages 545--555, 1972.

\bibitem[Ebegil et~al.(2021)Ebegil, {\"O}zdemir, and
  G{\"o}kpinar]{ebegil2021some}
Meral Ebegil, Yaprak~Arzu {\"O}zdemir, and Fikri G{\"o}kpinar.
\newblock Some shrinkage estimators based on median ranked set sampling.
\newblock \emph{Journal of Applied Statistics}, pages 1--26, 2021.

\bibitem[Felson et~al.(1993)Felson, Zhang, Hannan, and
  Anderson]{felson1993effects}
David~T Felson, Yuqing Zhang, Marian~T Hannan, and Jennifer~J Anderson.
\newblock Effects of weight and body mass index on bone mineral density in men
  and women: the framingham study.
\newblock \emph{Journal of Bone and Mineral Research}, 8\penalty0 (5):\penalty0
  567--573, 1993.

\bibitem[Frey(2012)]{frey2012nonparametric}
Jesse Frey.
\newblock Nonparametric mean estimation using partially ordered sets.
\newblock \emph{Environmental and ecological statistics}, 19\penalty0
  (3):\penalty0 309--326, 2012.

\bibitem[Hatefi and Alvandi(2020)]{hatefi2020efficient}
Armin Hatefi and Amirhossein Alvandi.
\newblock Efficient estimators with categorical ranked set samples: estimation
  procedures for osteoporosis.
\newblock \emph{Journal of Applied Statistics}, pages 1--16, 2020.

\bibitem[Hatefi and Jafari~Jozani(2017)]{hatefi2017improved}
Armin Hatefi and Mohammad Jafari~Jozani.
\newblock An improved procedure for estimation of malignant breast cancer
  prevalence using partially rank ordered set samples with multiple
  concomitants.
\newblock \emph{Statistical methods in medical research}, 26\penalty0
  (6):\penalty0 2552--2566, 2017.

\bibitem[Hatefi et~al.(2015)Hatefi, Jozani, and Ozturk]{hatefi2015mixture}
Armin Hatefi, Mohammad~Jafari Jozani, and Omer Ozturk.
\newblock Mixture model analysis of partially rank-ordered set samples: Age
  groups of fish from length-frequency data.
\newblock \emph{Scandinavian Journal of Statistics}, 42\penalty0 (3):\penalty0
  848--871, 2015.

\bibitem[Helu et~al.(2011)Helu, Samawi, and Vogel]{helu2011nonparametric}
Amal Helu, Hani Samawi, and Robert Vogel.
\newblock Nonparametric overlap coefficient estimation using ranked set
  sampling.
\newblock \emph{Journal of Nonparametric Statistics}, 23\penalty0 (2):\penalty0
  385--397, 2011.

\bibitem[Hoerl et~al.(1975)Hoerl, Kannard, and Baldwin]{hoerl1975ridge}
Arthur~E Hoerl, Robert~W Kannard, and Kent~F Baldwin.
\newblock Ridge regression: some simulations.
\newblock \emph{Communications in Statistics-Theory and Methods}, 4\penalty0
  (2):\penalty0 105--123, 1975.

\bibitem[Howard et~al.(1982)Howard, Jones, Mauldin, and
  Beal]{howard1982abundance}
Ralph~W Howard, Susan~C Jones, Joe~K Mauldin, and Raymond~H Beal.
\newblock Abundance, distribution, and colony size estimates for reticulitermes
  spp.(isoptera: Rhinotermitidae) in southern mississippi.
\newblock \emph{Environmental Entomology}, 11\penalty0 (6):\penalty0
  1290--1293, 1982.

\bibitem[Hubert and Wijekoon(2006)]{hubert2006improvement}
MH~Hubert and P~Wijekoon.
\newblock Improvement of the liu estimator in linear regression model.
\newblock \emph{Statistical Papers}, 47\penalty0 (3):\penalty0 471, 2006.

\bibitem[Inan and Erdogan(2013)]{inan2013liu}
Deniz Inan and Birsen~E Erdogan.
\newblock Liu-type logistic estimator.
\newblock \emph{Communications in Statistics-Simulation and Computation},
  42\penalty0 (7):\penalty0 1578--1586, 2013.

\bibitem[Kejian(1993)]{kejian1993new}
Liu Kejian.
\newblock A new class of blased estimate in linear regression.
\newblock \emph{Communications in Statistics-Theory and Methods}, 22\penalty0
  (2):\penalty0 393--402, 1993.

\bibitem[Kejian(2003)]{liu2003using}
Liu Kejian.
\newblock Using liu-type estimator to combat collinearity.
\newblock \emph{Communications in Statistics-Theory and Methods}, 32\penalty0
  (5):\penalty0 1009--1020, 2003.

\bibitem[Li and Yang(2010)]{li2010new}
Yalian Li and Hu~Yang.
\newblock A new stochastic mixed ridge estimator in linear regression model.
\newblock \emph{Statistical Papers}, 51\penalty0 (2):\penalty0 315--323, 2010.

\bibitem[Lim et~al.(2016)Lim, Kim, Lee, Byun, Park, and Kim]{lim2016comparison}
Hee-Sook Lim, Soon-Kyung Kim, Hae-Hyeog Lee, Dong~Won Byun, Yoon-Hyung Park,
  and Tae-Hee Kim.
\newblock Comparison in adherence to osteoporosis guidelines according to bone
  health status in korean adult.
\newblock \emph{Journal of bone metabolism}, 23\penalty0 (3):\penalty0
  143--148, 2016.

\bibitem[Lynne~Stokes(1977)]{lynne1977ranked}
S~Lynne~Stokes.
\newblock Ranked set sampling with concomitant variables.
\newblock \emph{Communications in Statistics-Theory and Methods}, 6\penalty0
  (12):\penalty0 1207--1211, 1977.

\bibitem[Melton~III et~al.(1998)Melton~III, Atkinson, O'connor, O'fallon, and
  Riggs]{melton1998bone}
L~Joseph Melton~III, Elizabeth~J Atkinson, Michael~K O'connor, W~Michael
  O'fallon, and B~Lawrence Riggs.
\newblock Bone density and fracture risk in men.
\newblock \emph{Journal of Bone and Mineral Research}, 13\penalty0
  (12):\penalty0 1915--1923, 1998.

\bibitem[Muttlak(1998)]{muttlak1998median}
HA~Muttlak.
\newblock Median ranked set sampling with concomitant variables and a
  comparison with ranked set sampling and regression estimators.
\newblock \emph{Environmetrics: The official journal of the International
  Environmetrics Society}, 9\penalty0 (3):\penalty0 255--267, 1998.

\bibitem[Muttlak(1995)]{muttlak1995parameters}
Hassen~A Muttlak.
\newblock Parameters estimation in a simple linear regression using rank set
  sampling.
\newblock \emph{Biometrical Journal}, 37\penalty0 (7):\penalty0 799--810, 1995.

\bibitem[Nahhas et~al.(2002)Nahhas, Wolfe, and Chen]{nahhas2002ranked}
Ramzi~W Nahhas, Douglas~A Wolfe, and Haiying Chen.
\newblock Ranked set sampling: cost and optimal set size.
\newblock \emph{Biometrics}, 58\penalty0 (4):\penalty0 964--971, 2002.

\bibitem[Neuburger et~al.(2015)Neuburger, Currie, Wakeman, Tsang, Plant,
  De~Stavola, Cromwell, and van~der Meulen]{neuburger2015impact}
Jenny Neuburger, Colin Currie, Robert Wakeman, Carmen Tsang, Fay Plant, Bianca
  De~Stavola, David~A Cromwell, and Jan van~der Meulen.
\newblock The impact of a national clinician-led audit initiative on care and
  mortality after hip fracture in england: an external evaluation using time
  trends in non-audit data.
\newblock \emph{Medical care}, 53\penalty0 (8):\penalty0 686, 2015.

\bibitem[Ozturk(2013)]{ozturk2013combining}
Omer Ozturk.
\newblock Combining multi-observer information in partially rank-ordered
  judgment post-stratified and ranked set samples.
\newblock \emph{Canadian Journal of Statistics}, 41\penalty0 (2):\penalty0
  304--324, 2013.

\bibitem[Ozturk(2014)]{ozturk2014estimation}
Omer Ozturk.
\newblock Estimation of population mean and total in a finite population
  setting using multiple auxiliary variables.
\newblock \emph{Journal of Agricultural, Biological, and Environmental
  Statistics}, 19\penalty0 (2):\penalty0 161--184, 2014.

\bibitem[Rao and Toutenburg(1995)]{rao1995linear}
Calyampudi~Radhakrishna Rao and Helge Toutenburg.
\newblock Linear models.
\newblock In \emph{Linear models}, pages 3--18. Springer, 1995.

\bibitem[Samawi and Al-Sagheer(2001)]{samawi2001estimation}
Hani~M Samawi and Omar~AM Al-Sagheer.
\newblock On the estimation of the distribution function using extreme and
  median ranked set sampling.
\newblock \emph{Biometrical Journal: Journal of Mathematical Methods in
  Biosciences}, 43\penalty0 (3):\penalty0 357--373, 2001.

\bibitem[Schaefer et~al.(1984)Schaefer, Roi, and Wolfe]{schaefer1984ridge}
RL~Schaefer, LD~Roi, and RA~Wolfe.
\newblock A ridge logistic estimator.
\newblock \emph{Communications in Statistics-Theory and Methods}, 13\penalty0
  (1):\penalty0 99--113, 1984.

\bibitem[Seeman et~al.(1983)Seeman, Melton~III, O'Fallon, and
  Riggs]{seeman1983risk}
Ego Seeman, L~Joseph Melton~III, W~Michael O'Fallon, and B~Lawrence Riggs.
\newblock Risk factors for spinal osteoporosis in men.
\newblock \emph{The American journal of medicine}, 75\penalty0 (6):\penalty0
  977--983, 1983.

\bibitem[Theil and Goldberger(1961)]{theil1961pure}
Henry Theil and Arthur~S Goldberger.
\newblock On pure and mixed statistical estimation in economics.
\newblock \emph{International Economic Review}, 2\penalty0 (1):\penalty0
  65--78, 1961.

\bibitem[Wang et~al.(2016)Wang, Lim, and Stokes]{wang2016using}
Xinlei Wang, Johan Lim, and Lynne Stokes.
\newblock Using ranked set sampling with cluster randomized designs for
  improved inference on treatment effects.
\newblock \emph{Journal of the American Statistical Association}, 111\penalty0
  (516):\penalty0 1576--1590, 2016.

\bibitem[Yang and Xu(2009)]{yang2009alternative}
Hu~Yang and Jianwen Xu.
\newblock An alternative stochastic restricted liu estimator in linear
  regression.
\newblock \emph{Statistical Papers}, 50\penalty0 (3):\penalty0 639--647, 2009.

\bibitem[Yang et~al.(2009)Yang, Chang, and Liu]{yang2009improvement}
Hu~Yang, Xinfeng Chang, and Deqiang Liu.
\newblock Improvement of the liu estimator in weighted mixed regression.
\newblock \emph{Communications in Statistics---Theory and Methods}, 38\penalty0
  (2):\penalty0 285--292, 2009.

\bibitem[Zamanzade and Wang(2018)]{zamanzade2018proportion}
Ehsan Zamanzade and Xinlei Wang.
\newblock Proportion estimation in ranked set sampling in the presence of tie
  information.
\newblock \emph{Computational Statistics}, 33\penalty0 (3):\penalty0
  1349--1366, 2018.

\end{thebibliography}
}

\newpage
\section{Appendix}

\subsection{ Proof of Lemma \ref{moments_ls_rss}}
\begin{align*}
\E({\widehat{\bf \beta}}_{\text{LS,MRS}}) = {\bf S}^{-1}_{\text{MRS}} {\bf X}_{\text{MRS}}  \E({\bf y}_{\text{MRS}} | {\bf X}_{\text{MRS}}) 
={\bf S}^{-1}_{\text{MRS}} {\bf X}_{\text{MRS}} {\bf X}^\top_{\text{MRS}} {\bf \beta} = \beta.
\end{align*}
\[
\text{Var}({\widehat{\bf \beta}}_{\text{LS,MRS}}) 
= \text{Var}({\bf S}^{-1}_{\text{MRS}} {\bf X}_{\text{MRS}} {\bf y}_{\text{MRS}}) 
= {\bf S}^{-1}_{\text{MRS}} {\bf X}_{\text{MRS}} \text{Var}({\bf y}_{\text{MRS}}|{\bf X}_{\text{MRS}}) {\bf X}^\top_{\text{MRS}} {\bf S}^{-1}_{\text{MRS}} 
= \sigma^2 {\bf S}^{-1}_{\text{MRS}}.  
\]
 The proof is completed by  the property of concomitant of order statistics where 
$\E({\bf y}_{\text{MRS}} | {\bf X}_{\text{MRS}}) = \E({\bf y} |{\bf X})$ and the fact that RSS data are independent statistics.
\hfill $\square$

\subsection{ Proof of Lemma \ref{moments_ridge_rss}} 
One can easily rewrite the ridge estimator  as follows
\begin{align} \label{ridge2}
{\widehat{\bf \beta}}_{\text{R,MRS}} = (\I + k {\bf S}^{-1}_{\text{MRS}})^{-1} {\widehat{\bf \beta}}_{\text{LS,MRS}}.
\end{align}
From  \eqref{ridge2} and Lemma \ref{moments_ls_rss}, one can easily complete the proof.  
 \hfill $\square$

\subsection{ Proof of Lemma \ref{moments_liu_rss}} 
We can rewrite the Liu estimator  as follows
\begin{align} \label{liu_ridge2} \nonumber
{\widehat{\bf \beta}}_{\text{LT,MRS}} &= 
({\bf S}_{\text{MRS}} + k \I)^{-1} ({\bf X}_{\text{MRS}}^\top 
{\bf y}_{\text{MRS}}+ d {\widehat{\bf \beta}}_{\text{R,MRS}}) \\\nonumber
&= ({\bf S}_{\text{MRS}} + k \I)^{-1} \left(\I + d ({\bf S}_{\text{MRS}} + k \I)^{-1} \right) {\bf S}_{\text{MRS}} 
{\widehat{\bf \beta}}_{\text{LS,MRS}}\\
&= {\bf A}_{\text{LT,MRS}} {\widehat{\bf \beta}}_{\text{LS,MRS}}.
\end{align}
Now from \eqref{liu_ridge2} and Lemma \ref{moments_ls_rss}, one can easily complete the proof.  
 \hfill $\square$

\subsection{ Proof of Lemma \ref{mixed_ls_rss}} 

\begin{itemize}
\item[i)] From restriction  model \ref{reg_sto} and $\text{Var}({\bf e})=\sigma^2 {\bf\Omega}$,  the penalized sum of least squares becomes   
\begin{align} \label{q_me}
{Q}_{\text{MRS}} =  ({\bf y}_{\text{MRS}} - {\bf X}_{\text{MRS}} {\bf \beta})^\top ({\bf y}_{\text{MRS}} - {\bf X}_{\text{MRS}} {\bf \beta})
+ v ({\bf\Omega}^{-1/2}{\bf r}- {\bf\Omega}^{-1/2} {\bf R} {\bf \beta})^\top ({\bf\Omega}^{-1/2}{\bf r}- {\bf\Omega}^{-1/2} {\bf R} {\bf \beta}).
\end{align}
From \eqref{q_me},  it is easy to  prove part (i). 
\item[ii)]
\begin{align*}
{\widehat{\bf \beta}}_{\text{ME,MRS}} 
&=  ({\bf S}_{\text{MRS}} + v  {\bf R}^\top {\bf \Omega}^{-1} {\bf R})^{-1} 
({\bf X}_{\text{MRS}}^\top {\bf y}_{\text{MRS}}+ v {\bf R}^\top {\bf \Omega}^{-1} {\bf r}) \\
&= ({\bf S}_{\text{MRS}} + v  {\bf R}^\top {\bf \Omega}^{-1} {\bf R})^{-1}{\bf X}_{\text{MRS}}^\top {\bf y}_{\text{MRS}} + 
({\bf S}_{\text{MRS}} + v  {\bf R}^\top {\bf \Omega}^{-1} v {\bf R}^\top {\bf \Omega}^{-1} {\bf r}) \\
&= \left({\bf S}_{\text{MRS}}^{-1} + v {\bf S}_{\text{MRS}}^{-1} {\bf R}^\top ({\bf \Omega} + v {\bf R} {\bf \Omega}^{-1} {\bf R}^\top)^{-1}
 {\bf R}{\bf S}_{\text{MRS}}^{-1} \right) {\bf X}_{\text{MRS}}^\top {\bf y}_{\text{MRS}}\\
&~~~~+ v {\bf S}_{\text{MRS}}^{-1} {\bf R}^\top ({\bf \Omega} + v {\bf R} {\bf \Omega}^{-1} {\bf R}^\top)^{-1} {\bf r} \\
&= {\bf S}_{\text{MRS}}^{-1} {\bf X}_{\text{MRS}}^\top {\bf y}_{\text{MRS}} -
   v {\bf S}_{\text{MRS}}^{-1} {\bf R}^\top ({\bf \Omega} + v {\bf R} {\bf \Omega}^{-1} {\bf R}^\top)^{-1} {\bf R} {\bf S}_{\text{MRS}}^{-1}
   {\bf X}_{\text{MRS}}^\top {\bf y}_{\text{MRS}} \\
&~~~~+ v {\bf S}_{\text{MRS}}^{-1} {\bf R}^\top ({\bf \Omega} + v {\bf R} {\bf \Omega}^{-1} {\bf R}^\top)^{-1} {\bf r} \\
&=  {\widehat{\bf \beta}}_{\text{LS,MRS}} + 
 v {\bf S}_{\text{MRS}}^{-1} {\bf R}^\top ({\bf \Omega} + v {\bf R} {\bf \Omega}^{-1} {\bf R}^\top)^{-1} 
 \left( {\bf r} -  {\bf R} {\bf S}_{\text{MRS}}^{-1} {\bf X}_{\text{MRS}}^\top {\bf y}_{\text{MRS}} \right) \\
&=  {\widehat{\bf \beta}}_{\text{LS,MRS}} + 
 v {\bf S}_{\text{MRS}}^{-1} {\bf R}^\top ({\bf \Omega} + v {\bf R} {\bf \Omega}^{-1} {\bf R}^\top)^{-1} 
 \left( {\bf r} -  {\bf R} {\widehat{\bf \beta}}_{\text{LS,MRS}} \right), 
\end{align*} 
where the third equality is implied by Lemma \ref{lemma_rao}. \hfill $\square$
\end{itemize}

\subsection{ Proof of Lemma \ref{moments_srl_rss}} 

Let ${\bf F}_{{\text{MRS}},d} = ({\bf S}_{\text{MRS}} +\I)^{-1} ({\bf S}_{\text{MRS}} +d \I)$. We first show
\begin{align} \label{le1_2} \nonumber
{\widehat{\bf \beta}}_{\text{LT1,MRS}} &= 
({\bf S}_{\text{MRS}} + \I)^{-1} ({\bf X}_{\text{MRS}}^\top {\bf y}_{\text{MRS}} + d {\widehat{\bf \beta}}_{\text{LS,MRS}}) \\\nonumber
&= ({\bf S}_{\text{MRS}} + \I)^{-1} ({\bf S}_{\text{MRS}} {\widehat{\bf \beta}}_{\text{LS,MRS}} + d {\widehat{\bf \beta}}_{\text{LS,MRS}}) \\
&= {\bf F}_{{\text{MRS}},d} {\bf S}_{\text{MRS}}^{-1} {\bf X}_{\text{MRS}}^\top {\bf y}_{\text{MRS}}.
\end{align}
From \eqref{le1_2} and \eqref{srr_rss} and that ${\bf F}_{{\text{MRS}},d}$ and  ${\bf S}_{\text{MRS}}^{-1}$ are commutative,  we derive
 ${\widehat{\bf \beta}}_{\text{SRL,MRS}}$ as follows
\begin{align} \label{srl_mrs_2} \nonumber
{\widehat{\bf \beta}}_{\text{SRL,MRS}} &= 
{\bf S}_{\text{MRS}}^{-1} {\bf F}_{{\text{MRS}},d}  {\bf X}_{\text{MRS}}^\top {\bf y}_{\text{MRS}} + 
v {\bf S}_{\text{MMR}}^{-1} {\bf R}^\top ({\bf \Omega} + v {\bf R} {\bf S}_{\text{MMR}}^{-1} {\bf R}^\top)^{-1} 
({\bf r}- {\bf R} {\bf S}_{\text{MRS}}^{-1} {\bf F}_{{\text{MRS}},d}  {\bf X}_{\text{MRS}}^\top {\bf y}_{\text{MRS}} ) \\\nonumber
&= {\bf S}_{\text{MRS}}^{-1} {\bf F}_{{\text{MRS}},d}  {\bf X}_{\text{MRS}}^\top {\bf y}_{\text{MRS}} + 
v {\bf S}_{\text{MMR}}^{-1} {\bf R}^\top ({\bf \Omega} + v {\bf R} {\bf S}_{\text{MMR}}^{-1} {\bf R}^\top)^{-1} {\bf r} \\\nonumber
&~~ - v {\bf S}_{\text{MMR}}^{-1} {\bf R}^\top ({\bf \Omega} + v {\bf R} {\bf S}_{\text{MMR}}^{-1} {\bf R}^\top)^{-1} {\bf R} {\bf S}_{\text{MRS}}^{-1} {\bf F}_{{\text{MRS}},d}  {\bf X}_{\text{MRS}}^\top {\bf y}_{\text{MRS}} \\\nonumber
&= \left( {\bf S}_{\text{MRS}}^{-1} - v {\bf S}_{\text{MMR}}^{-1} {\bf R}^\top 
({\bf \Omega} + v {\bf R} {\bf S}_{\text{MMR}}^{-1} {\bf R}^\top)^{-1} {\bf R} {\bf S}_{\text{MRS}}^{-1} \right) 
\left( {\bf F}_{{\text{MRS}},d}  {\bf X}_{\text{MRS}}^\top {\bf y}_{\text{MRS}} + v {\bf R}^\top {\bf \Omega}^{-1} {\bf r} \right) \\
&= ( {\bf S}_{\text{MRS}} + v {\bf R}^\top {\bf \Omega}^{-1} {\bf R} )^{-1} 
({\bf F}_{{\text{MRS}},d}  {\bf X}_{\text{MRS}}^\top {\bf y}_{\text{MRS}}  + v {\bf R}^\top {\bf \Omega}^{-1} {\bf r}). 
\end{align}
From \eqref{srl_mrs_2} and Lemma \ref{moments_ls_rss}, the expected value of ${\widehat{\bf \beta}}_{\text{SRL,MRS}}$ yields
\begin{align*}
\E({\widehat{\bf \beta}}_{\text{SRL,MRS}}) 
&=  ( {\bf S}_{\text{MRS}} + v {\bf R}^\top {\bf \Omega}^{-1} {\bf R} )^{-1} 
{\bf F}_{{\text{MRS}},d}  {\bf X}_{\text{MRS}}^\top \E({\bf y}_{\text{MRS}}| {\bf X}_{\text{MRS}})  +
( {\bf S}_{\text{MRS}} + v {\bf R}^\top {\bf \Omega}^{-1} {\bf R} )^{-1}   v {\bf R}^\top {\bf \Omega}^{-1} \E({\bf r}) \\
&= {\bf \beta} + ( {\bf S}_{\text{MRS}} + v {\bf R}^\top {\bf \Omega}^{-1} {\bf R} )^{-1}
 ({\bf F}_{{\text{MRS}},d} - \I) {\bf S}_{\text{MRS}} {\bf \beta}.
\end{align*}
\begin{align*}
\text{Var}({\widehat{\bf \beta}}_{\text{SRL,MRS}}) 
&= {\bf A}_{\text{SRL,MRS}} {\bf F}_{{\text{MRS}},d} {\bf X}_{\text{MRS}}^\top  \text{Var}({\bf y}_{\text{MRS}}| {\bf X}_{\text{MRS}})
 {\bf X}_{\text{MRS}} {\bf F}_{{\text{MRS}},d}^\top {\bf A}_{\text{SRL,MRS}}   \\
&~~ +  v^2 {\bf A}_{\text{SRL,MRS}} {\bf R}^\top {\bf \Omega}^{-1} \text{Var}({\bf r}) {\bf \Omega}^{-1} {\bf R} {\bf A}_{\text{SRL,MRS}} \\
&=  \sigma^2 {\bf A}_{\text{SRL,MRS}} \left( {\bf F}_{{\text{MRS}},d} {\bf S}_{\text{MRS}} {\bf F}_{{\text{MRS}},d}^\top + 
v^2 {\bf R}^\top {\bf \Omega}^{-1} {\bf R} \right) {\bf A}_{\text{SRL,MRS}},
\end{align*} 
where ${\bf A}_{\text{SRL,MRS}}=({\bf S}_{\text{MRS}} + v {\bf R}^\top {\bf \Omega}^{-1} {\bf R})^{-1}$ and the first equality is implied by \eqref{srl_mrs_2} and Lemma \ref{moments_ls_rss}. \hfill $\square$

\subsection{ Proof of Lemma \ref{moments_srr_rss}} 
Let ${\bf W}_{\text{MRS}} = (\I +k {\bf S}_{\text{MRS}}^{-1})^{-1}$. From \eqref{ridge2} and that fact that ${\bf W}_{\text{MRS}}$  and  ${\bf S}_{\text{MRS}}^{-1}$ are commutative,  we show
 \begin{align}\label{srr_mrs_2} \nonumber
{\widehat{\bf \beta}}_{\text{SRR,MRS}} &= 
{\bf W}_{\text{MRS}}  {\widehat{\bf \beta}}_{\text{LS,MRS}}  
+ v {\bf S}_{\text{MRS}}^{-1} {\bf R}^\top ({\bf \Omega} + v {\bf R} {\bf S}_{\text{MRS}}^{-1} {\bf R}^\top)^{-1}
 ({\bf r}- {\bf R} {\bf W}_{\text{MRS}}  {\widehat{\bf \beta}}_{\text{LS,MRS}}) \\\nonumber
 &= {\bf S}_{\text{MRS}}^{-1} {\bf W}_{\text{MRS}}  {\bf X}_{\text{MRS}}^\top {\bf y}_{\text{MRS}} 
 + v {\bf S}_{\text{MRS}}^{-1} {\bf R}^\top ({\bf \Omega} + v {\bf R} {\bf S}_{\text{MRS}}^{-1} {\bf R}^\top)^{-1}
 ({\bf r}- {\bf R} {\bf S}_{\text{MRS}}^{-1} {\bf W}_{\text{MRS}}  {\bf X}_{\text{MRS}}^\top {\bf y}_{\text{MRS}}) \\\nonumber
 &= \left( {\bf S}_{\text{MRS}}^{-1} - v {\bf S}_{\text{MMR}}^{-1} {\bf R}^\top 
({\bf \Omega} + v {\bf R} {\bf S}_{\text{MMR}}^{-1} {\bf R}^\top)^{-1} {\bf R} {\bf S}_{\text{MRS}}^{-1} \right) 
\left( {\bf W}_{\text{MRS}}  {\bf X}_{\text{MRS}}^\top {\bf y}_{\text{MRS}} + v {\bf R}^\top {\bf \Omega}^{-1} {\bf r}\right)\\
&= ({\bf S}_{\text{MRS}} + v {\bf R}^\top {\bf \Omega}^{-1} {\bf R})^{-1} \left( {\bf W}_{\text{MRS}}  {\bf X}_{\text{MRS}}^\top {\bf y}_{\text{MRS}} +
 v{\bf R}^\top {\bf \Omega}^{-1} {\bf r}\right).
\end{align}
Using \ref{srr_mrs_2} and Lemma \ref{moments_ls_rss}, we compute the expected value of ${\widehat{\bf \beta}}_{\text{SRR,MRS}}$ as
\begin{align*}
\E({\widehat{\bf \beta}}_{\text{SRR,MRS}}) 
&=  ({\bf S}_{\text{MRS}} + v {\bf R}^\top {\bf \Omega}^{-1} {\bf R})^{-1}
{\bf W}_{\text{MRS}}  {\bf X}_{\text{MRS}}^\top \E({\bf y}_{\text{MRS}}| {\bf X}_{\text{MRS}})  +
({\bf S}_{\text{MRS}} + v {\bf R}^\top {\bf \Omega}^{-1} {\bf R})^{-1}   v{\bf R}^\top {\bf \Omega}^{-1} \E({\bf r}) \\
&= ({\bf S}_{\text{MRS}} + v {\bf R}^\top {\bf \Omega}^{-1} {\bf R})^{-1}
{\bf W}_{\text{MRS}}  {\bf S}_{\text{MRS}} {\bf \beta} + 
({\bf S}_{\text{MRS}} + v {\bf R}^\top {\bf \Omega}^{-1} {\bf R})^{-1} 
v {\bf R}^\top {\bf \Omega}^{-1} {\bf R})^{-1}   v{\bf R}^\top {\bf \Omega}^{-1} {\bf R} {\bf \beta} \\
&= {\bf \beta} + ({\bf S}_{\text{MRS}} + v {\bf R}^\top {\bf \Omega}^{-1} {\bf R})^{-1} 
\left( {\bf W}_{\text{MRS}} - \I \right) {\bf S}_{\text{MRS}} {\bf \beta}.
\end{align*}
\begin{align*}
\text{Var}({\widehat{\bf \beta}}_{\text{SRR,MRS}})
&=  {\bf A}_{\text{SRR,MSR}} {\bf W}_{\text{MRS}}  {\bf X}_{\text{MRS}}^\top \text{Var}({\bf y}_{\text{MRS}}| {\bf X}_{\text{MRS}})
{\bf X}_{\text{MRS}}  {\bf W}_{\text{MRS}}^\top {\bf A}_{\text{SRR,MSR}} \\
&~~ +  v^2 {\bf A}_{\text{SRR,MSR}} {\bf R}^\top {\bf \Omega}^{-1} \text{Var}({\bf r}) {\bf \Omega}^{-1} {\bf R} {\bf A}_{\text{SRR,MSR}}  \\
&= \sigma^2  {\bf A}_{\text{SRR,MSR}} \left( {\bf W}_{\text{MRS}} {\bf S}_{\text{MRS}} {\bf W}_{\text{MRS}}^\top + 
v^2 {\bf R}^\top {\bf \Omega}^{-1} {\bf R} \right) {\bf A}_{\text{SRR,MSR}},
\end{align*}
where ${\bf A}_{\text{SRR,MSR}}= ({\bf S}_{\text{MRS}} + v {\bf R}^\top {\bf \Omega}^{-1} {\bf R})^{-1}$ and the first equality is implied by \eqref{srr_mrs_2} and Lemma \ref{moments_ls_rss}. \hfill $\square$

\newpage


\newpage

\begin{table}[ht]
\centering
\footnotesize{\begin{tabular}{c|c|cccccc}
  \hline
$n$ & $\kappa$ &  ${\widehat{\bf \beta}}_{\text{SMRE,SRS}}$ &  ${\widehat{\bf \beta}}_{\text{SMRE,MMRS}}$
    &  ${\widehat{\bf \beta}}_{\text{SMRE,MMRM}}$ &  ${\widehat{\bf \beta}}_{\text{SLE,SRS}}$
     &  ${\widehat{\bf \beta}}_{\text{SLE,MMRS}}$ &  ${\widehat{\bf \beta}}_{\text{SLE,MMRM}}$ \\ 
  \hline
   3& 0.75 & 0.493 & 0.342 & 0.339 & 1.289 & 0.723 & 0.721 \\ [-1ex]
   & 0.80 & 0.334 & 0.263 & 0.262 & 1.284 & 0.691 & 0.682 \\ [-1ex]
   & 0.85 & 0.277 & 0.247 & 0.247 & 1.245 & 0.648 & 0.663 \\ [-1ex]
   & 0.90 & 0.254 & 0.246 & 0.245 & 1.248 & 0.600 & 0.580 \\ [-1ex]
   & 0.95 & 0.255 & 0.250 & 0.251 & 14.079 & 2.923 & 17.492 \\ [-1ex]
   & 0.99 & 0.259 & 0.256 & 0.256 & 0.296 & 0.441 & 0.321 \\ 
  \cline{1-8}
   4& 0.75 & 0.439 & 0.304 & 0.308 & 1.235 & 0.692 & 0.695 \\ [-1ex]
   & 0.80 & 0.308 & 0.243 & 0.243 & 1.269 & 0.667 & 0.656 \\ [-1ex]
   & 0.85 & 0.264 & 0.239 & 0.239 & 1.246 & 0.624 & 0.626 \\ [-1ex]
   & 0.90 & 0.251 & 0.244 & 0.244 & 1.109 & 0.569 & 0.566 \\ [-1ex]
   & 0.95 & 0.253 & 0.250 & 0.250 & 14.108 & 163 & 63.751 \\ [-1ex]
   & 0.99 & 0.257 & 0.255 & 0.255 & 0.434 & 11.426 & 0.623 \\ 
   \hline
\end{tabular}
\caption{\footnotesize{The REs of ${\widehat{\bf\beta}}_{\text{SMRE}}$, ${\widehat{\bf\beta}}_{\text{SLE}}$ under SRS, MRS and MMR data relative to their
${\widehat{\bf\beta}}_{\text{LS,SRS}}$ counterpart of the same size when  $H= 4$ and $c=1$ with ranking ability $\rho=(1,1,1)$.}}
\label{tab_sim_sto_H4}}
\end{table}

\newpage

\begin{table}[ht]
    \centering
   \footnotesize{ \begin{tabular}{c|c|cccc}
      \hline
      $\kappa$ & $c$ & ${\widehat{\bf\beta}}_{\text{SMRE,SRS}}$ & ${\widehat{\bf\beta}}_{\text{SMRE,MMR}}$ & 
      ${\widehat{\bf\beta}}_{\text{SLE,SRS}}$ & ${\widehat{\bf\beta}}_{\text{SLE,MMR}}$ \\ 
      \hline
       0.75&0.1 & 0.48638 & 0.31411 & 1.27164 & 0.63137 \\ [-1ex]
           &0.5 & 0.49507 & 0.32370 & 1.28471 & 0.65129 \\ [-1ex]
           &1.0 & 0.49255 & 0.33905 & 1.28901 & 0.72098 \\ 
     \hline
        0.8&0.1 & 0.33630 & 0.24889 & 1.29035 & 0.59610 \\ [-1ex]
           &0.5 & 0.33329 & 0.25168 & 1.26088 & 0.62620 \\ [-1ex]
           &1 & 0.33432 & 0.26249 & 1.28387 & 0.68222 \\ 
      \hline
       0.85&0.1 & 0.27714 & 0.24186 & 1.21881 & 0.55860 \\ [-1ex]
           &0.5 & 0.27704 & 0.24341 & 1.24289 & 0.58399 \\ [-1ex]
           &1 & 0.27666 & 0.24701 & 1.24523 & 0.66252 \\ 
     \hline
      0.9 &0.1 & 0.25451 & 0.24411 & 1.08373 & 5.34304 \\ [-1ex]
          &0.5 & 0.25400 & 0.24514 & 1.64597 & 0.69039 \\ [-1ex]
          &1 & 0.25430 & 0.24524 & 1.24823 & 0.58050 \\ 
      \hline
     0.95 &0.1 & 0.25421 & 0.24995 & 59.38280 & 558.768 \\ [-1ex]
          &0.5 & 0.25493 & 0.25037 & 36.05183 & 13.2592 \\ [-1ex]
          &1 & 0.25478 & 0.25094 & 14.07878 & 17.4918 \\ 
      \hline
      0.99&0.1 & 0.25864 & 0.25519 & 0.56044 & 0.55860 \\ [-1ex]
          &0.5 & 0.25869 & 0.25536 & 7.37324 & 0.58399 \\ [-1ex]
           &1 & 0.25929 & 0.25602 & 0.29642 & 0.66252 \\ 
      \hline
    \end{tabular}
    \caption{\footnotesize{The REs of ${\widehat{\bf\beta}}_{\text{SMRE}}$, ${\widehat{\bf\beta}}_{\text{SLE}}$ under SRS and MMR data relative to their
${\widehat{\bf\beta}}_{\text{LS,SRS}}$ counterpart of the same size with $N=12$, $H =4$  with ranking ability $\rho=(1,1,1)$.}}
    \label{tab_sim_sto_c_r}}
\end{table}


\newpage

\begin{table}[ht]
\centering
\footnotesize{\begin{tabular}{c|c|ccc|ccc}
  \hline
    &  & &$n=4, H=6$ &  & &$n=2, H=12$ &  \\
  \cline{3-8}
  $\phi$& Estimator  & Median & 2.5\% & 97.5\% & Median & 2.5\% & 97.5\% \\ 
  \hline
0.95& ${\widehat{\bf\beta}}_{\text{LS,SRS}}$ & 8146.35 & 350.71 & 141588 & 8335.14 & 330.48 & 137471 \\ [-1ex]
    & ${\widehat{\bf\beta}}_{\text{R,SRS}}$  & 478.35  & 42.25  & 7093   & 492.66  & 37.17  & 7163 \\ [-1ex]
    & ${\widehat{\bf\beta}}_{\text{LT,SRS}}$ & 76.23   & 12.82  & 1083   & 78.70   & 12.40  & 1058 \\ 
    & ${\widehat{\bf\beta}}_{\text{LS,RSS}}$ & 8018.13 & 314.70 & 113177 & 7781.22 & 346.63 & 119211 \\ [-1ex]
    & ${\widehat{\bf\beta}}_{\text{R,RSS}}$  & 482.96  & 41.18  & 6417   & 466.35  & 42.86  & 6490 \\ [-1ex]
    & ${\widehat{\bf\beta}}_{\text{LT,RSS}}$ & 77.57   & 12.65  & 1015   & 74.71   & 13.11  & 1072 \\ 
    & ${\widehat{\bf\beta}}_{\text{LS,MRS}}$ & 7749.11 & 356.69 & 104580 & 7269.18 & 308.65 & 104153 \\ [-1ex]
    & ${\widehat{\bf\beta}}_{\text{R,MRS}}$  & 474.22  & 41.78  & 5760   & 448.40  & 41.62  & 5998 \\ [-1ex]
    & ${\widehat{\bf\beta}}_{\text{LT,MRS}}$ & 76.74   & 13.18  & 997    & 72.81   & 12.72 & 1009 \\ 
  \hline
0.98& ${\widehat{\bf\beta}}_{\text{LS,SRS}}$ & 12316.19 & 565.68 & 158677 & 12137.58 & 570.44 & 150771 \\ [-1ex]
    & ${\widehat{\bf\beta}}_{\text{R,SRS}}$  & 364.46   & 45.09  & 6921   & 341.25   & 42.22  & 6693 \\ [-1ex]
    & ${\widehat{\bf\beta}}_{\text{LT,SRS}}$ & 58.22    & 13.95  & 808    & 56.22    & 13.09  & 793 \\ 
    & ${\widehat{\bf\beta}}_{\text{LS,RSS}}$ & 12252.13 & 559.89 & 142438 & 12145.12 & 580.39 & 133728 \\ [-1ex]
    & ${\widehat{\bf\beta}}_{\text{R,RSS}}$  & 350.58   & 44.49  & 6747   & 341.58   & 45.47  & 6232 \\ [-1ex]
    & ${\widehat{\bf\beta}}_{\text{LT,RSS}}$ & 56.44    & 13.92  & 817    & 56.12    & 13.93  & 731 \\    
    & ${\widehat{\bf\beta}}_{\text{LS,MRS}}$ & 11925.81 & 593.71 & 129248 & 11453.22 & 582.81 & 124213 \\ [-1ex]
    & ${\widehat{\bf\beta}}_{\text{R,MRS}}$  & 349.00   & 42.52  & 6485   & 336.47   & 43.31  & 6525 \\ [-1ex]
    & ${\widehat{\bf\beta}}_{\text{LT,MRS}}$ & 55.47    & 13.49  & 723    & 54.69    & 13.69  & 751 \\ 
   \hline
\end{tabular}
\caption{\footnotesize{The median and 95\% CI for the MSE of estimators of coefficients of the logistic model 
under SRS, RSS, MRS and MMR data of the same size when $\eta=0.98$, $c=0.2$ and ranking ability $\rho=(0.95,0.95)$.}}
\label{tab_sim_log_r98_good}}
\end{table}

\newpage

\begin{table}[ht]
\centering
\footnotesize{\begin{tabular}{c|c|ccc|ccc}
  \hline
    &  & &$n=4, H=6$ &  & &$n=2, H=12$ &  \\
  \cline{3-8}
  $\phi$& Estimator  & Median & 2.5\% & 97.5\% & Median & 2.5\% & 97.5\% \\ 
  \hline
0.95& ${\widehat{\bf\beta}}_{\text{LS,SRS}}$ & 2635.56 & 237.25 & 23294 & 2660.21 & 246.47 & 24389 \\ [-1ex]
    & ${\widehat{\bf\beta}}_{\text{R,SRS}}$  & 299.16  & 38.18 & 3802 & 294.72 & 39.02 & 3396 \\ [-1ex]
    & ${\widehat{\bf\beta}}_{\text{LT,SRS}}$ & 58.17   & 12.06 & 670 & 56.79 & 12.15 & 710 \\ 
    & ${\widehat{\bf\beta}}_{\text{LS,RSS}}$ & 2635.15 & 238.40 & 20987 & 2480.01 & 235.60 & 21816 \\ [-1ex]
    & ${\widehat{\bf\beta}}_{\text{R,RSS}}$  & 309.76  & 37.36 & 3616 & 284.17 & 37.01 & 3946 \\ [-1ex]
    & ${\widehat{\bf\beta}}_{\text{LT,RSS}}$ & 59.91   & 12.02 & 706 & 57.44 & 12.03 & 688 \\ 
    & ${\widehat{\bf\beta}}_{\text{LS,MRS}}$ & 2503.00 & 245.12 & 21639 & 2468.72 & 235.50 & 23036 \\ [-1ex]
    & ${\widehat{\bf\beta}}_{\text{R,MRS}}$  & 293.57  & 38.41 & 3101 & 280.75 & 37.58 & 4192 \\ [-1ex]
    & ${\widehat{\bf\beta}}_{\text{LT,MRS}}$ & 57.30   & 11.97 & 632 & 54.59 & 12.10 & 785 \\ 
  \hline
0.98& ${\widehat{\bf\beta}}_{\text{LS,SRS}}$ & 4410.32 & 327.17 & 55437 & 4214.86 & 309.32 & 50111 \\ [-1ex]
    & ${\widehat{\bf\beta}}_{\text{R,SRS}}$  & 324.51  & 38.93 & 3423 & 303.68 & 39.87 & 3315 \\ [-1ex]
    & ${\widehat{\bf\beta}}_{\text{LT,SRS}}$ & 56.98   & 12.89 & 657 & 53.69 & 13.09 & 631 \\ 
    & ${\widehat{\bf\beta}}_{\text{LS,RSS}}$ & 4354.90 & 340.09 & 43784 & 4298.78 & 353.43 & 46839 \\ [-1ex]
    & ${\widehat{\bf\beta}}_{\text{R,RSS}}$  & 321.45  & 42.36 & 3285 & 313.95 & 43.35 & 3343 \\ [-1ex]
    & ${\widehat{\bf\beta}}_{\text{LT,RSS}}$ & 57.78   & 13.34 & 605 & 55.79 & 13.35 & 668 \\ 
    & ${\widehat{\bf\beta}}_{\text{LS,MRS}}$ & 4240.66 & 295.37 & 45573 & 4386.70 & 328.58 & 46921 \\ [-1ex]
    & ${\widehat{\bf\beta}}_{\text{R,MRS}}$  & 307.44  & 37.60 & 3429 & 310.22 & 40.33 & 3313 \\ [-1ex]
    & ${\widehat{\bf\beta}}_{\text{LT,MRS}}$ & 54.32   & 12.23 & 659 & 55.08 & 12.95 & 633\\ 
   \hline
\end{tabular}
\caption{\footnotesize{The median and 95\% CI for the MSE of estimators of coefficients of the logistic model 
under SRS, RSS, MRS and MMR data of the same size when $\eta=0.95$, $c=0.2$ and ranking ability $\rho=(0.5,0.5)$.}}
\label{tab_sim_log_r95_bad}}
\end{table}

\newpage

\begin{table}[ht]
\centering
\footnotesize{\begin{tabular}{c|c|ccc|ccc}
  \hline
    &  & &$n=4, H=6$ &  & &$n=2, H=12$ &  \\
  \cline{3-8}
  $\phi$& Estimator  & Median & 2.5\% & 97.5\% & Median & 2.5\% & 97.5\% \\ 
  \hline
0.95& ${\widehat{\bf\beta}}_{\text{LS,SRS}}$ & 8186.92 & 322.25 & 147400 & 8200.18 & 284.92 & 128528 \\ [-1ex]
    & ${\widehat{\bf\beta}}_{\text{R,SRS}}$  & 472.71  & 41.20  & 7086   & 470.02 & 38.31 & 6824 \\ [-1ex]
    & ${\widehat{\bf\beta}}_{\text{LT,SRS}}$ & 76.43   & 13.00  & 1111   & 73.79 & 12.58 & 1082 \\ 
    & ${\widehat{\bf\beta}}_{\text{LS,RSS}}$ & 8106.87 & 304.24 & 116801 & 8372.63 & 372.34 & 114303 \\ [-1ex]
    & ${\widehat{\bf\beta}}_{\text{R,RSS}}$  & 467.80  & 39.51  & 6718   & 503.27 & 42.88 & 6466 \\ [-1ex]
    & ${\widehat{\bf\beta}}_{\text{LT,RSS}}$ & 76.01   & 12.91  & 1002   & 80.10 & 13.26 & 1119 \\    
    & ${\widehat{\bf\beta}}_{\text{LS,MRS}}$ & 7822.96 & 331.31 & 112281 & 7956.18 & 295.11 & 120813 \\ [-1ex]
    & ${\widehat{\bf\beta}}_{\text{R,MRS}}$  & 459.36  & 42.74  & 6587   & 476.00 & 39.83 & 6582 \\ [-1ex]
    & ${\widehat{\bf\beta}}_{\text{LT,MRS}}$ & 73.70   & 13.08  & 1056   & 76.28 & 12.58 & 1058 \\ 
  \hline
0.98& ${\widehat{\bf\beta}}_{\text{LS,SRS}}$ & 12461.70 & 573.95 & 152708 & 12285.95 & 535.29 & 164743 \\ [-1ex]
    & ${\widehat{\bf\beta}}_{\text{R,SRS}}$  & 360.71   & 45.37  & 6604 & 334.03 & 41.14 & 7232 \\ [-1ex]
    & ${\widehat{\bf\beta}}_{\text{LT,SRS}}$ & 58.11    & 13.86  & 758 & 55.56 & 13.04 & 866 \\ 
    & ${\widehat{\bf\beta}}_{\text{LS,RSS}}$ & 12874.86 & 685.43 & 146421 & 12784.51 & 535.32 & 146359 \\ [-1ex]
    & ${\widehat{\bf\beta}}_{\text{R,RSS}}$  & 372.82   & 46.75  & 6818 & 376.91 & 43.59 & 6699 \\ [-1ex]
    & ${\widehat{\bf\beta}}_{\text{LT,RSS}}$ & 59.06    & 13.92  & 779 & 59.20 & 13.61 & 806 \\    
    & ${\widehat{\bf\beta}}_{\text{LS,MRS}}$ & 12198.05 & 550.33 & 129290 & 11936.86 & 521.83 & 141357 \\ [-1ex]
    & ${\widehat{\bf\beta}}_{\text{R,MRS}}$  & 361.00   & 42.29  & 6790 & 352.16 & 41.51 & 6422 \\ [-1ex]
    & ${\widehat{\bf\beta}}_{\text{LT,MRS}}$ & 57.81    & 13.10  & 773 & 56.68 & 13.33 & 725 \\ 
   \hline
\end{tabular}
\caption{\footnotesize{The median and 95\% CI for the MSE of estimators of coefficients of the logistic model 
under SRS, RSS, MRS and MMR data of the same size when $\eta=0.98$, $c=0.2$ and ranking ability $\rho=(0.5,0.5)$.}}
\label{tab_sim_log_r98_bad}}
\end{table}

\newpage

\newpage

\newpage

\newpage

\end{document}